\documentclass{elsart}

\usepackage{indentfirst}

\usepackage[latin1]{inputenc}
\usepackage{graphicx}
\usepackage{subfigure}
\usepackage{psfrag}

\usepackage{amsmath, amssymb}

\usepackage{xspace} 

\usepackage{color}
\definecolor{oneblue}{rgb}{0,0.0,0.75}

\newtheorem{proposition}{Proposition}

\newtheorem{remark}{Remark}

\newcommand{\od}[2]{\frac{d#1}{d#2}}
\newcommand{\pd}[2]{\frac{\partial#1}{\partial#2}}
\newcommand{\set}[1]{\left\{ #1 \right\}}
\newcommand{\vol}{\mathop{\mathrm{vol}}}
\newcommand{\area}{\mathop{\mathrm{area}}}

\def\g{\vec{g}}
\def\u{\vec{u}}

\def\vv{\mathbf{v}}
\def\div{\nabla\cdot}
\def\I{\mathbb{I}}
\def\w{\mathbf{w}}
\def\F{\mathcal{F}}
\def\S{\mathcal{S}}

\def\n{\vec{n}}
\def\A{\mathbb{A}}

\def\N{\mathcal{N}}

\def\x{\vec{x}}

\def\R{\mathbb{R}}
\def\rit{\mathbb{R}}

\def\T{\mathcal{T}}

\def\sgn{\mbox{sgn}}

\journal{Computers and Fluids}

\begin{document}

\begin{frontmatter}

\title{A two-fluid model for violent aerated flows}

\author{\corauthref{cor}Fr\'{e}d\'{e}ric Dias}\ead{Frederic.Dias@cmla.ens-cachan.fr},\,
\author{Denys Dutykh$^1$}\ead{Denys.Dutykh@cmla.ens-cachan.fr}\,
and \author{Jean-Michel Ghidaglia}\ead{jmg@cmla.ens-cachan.fr}
\address{Centre de Mathématiques et de Leurs Applications,\\ ENS Cachan and CNRS, UniverSud, 61 avenue du
  President Wilson,\\  F-94235 Cachan Cedex, and LRC MESO, ENS Cachan, CEA DAM DIF\\
  $^1$ Now at Université de Savoie, Laboratoire de Mathématiques LAMA - UMR 5127\\
Campus Scientifique, 73376 Le Bourget-du-Lac Cedex.}
\corauth[cor]{Corresponding author.}

\begin{abstract}
In the study of ocean wave impact on structures, one often uses Froude scaling since the dominant force is gravity.
However the presence of trapped or entrained air in the water can significantly modify wave impacts. When air is
entrained in water in the form of small bubbles, the acoustic properties in the water change dramatically. While some
work has been done to study small-amplitude disturbances in such mixtures, little work has been done on large
disturbances in air-water mixtures. We propose a basic two-fluid model in which both fluids share the same velocities
and analyze some of its properties.
It is shown that this model can successfully mimic water wave impacts on coastal structures. The governing equations are discretized by a second-order finite volume method. Numerical results are presented for two examples: the dam break problem
and the drop test problem. The results suggest that this basic model can be used to study violent aerated flows, especially by providing fast qualitative estimates.
\end{abstract}


\end{frontmatter}


\section{Introduction}

One of the challenges in Computational Fluid Dynamics (CFD) is to determine efforts exerted by waves on structures,
especially coastal structures. The flows associated with wave impact can be quite complicated. In particular, wave breaking can lead to flows that cannot be described by models like {\it e.g.} the free-surface Euler or Navier--Stokes equations.
In a free-surface model, the boundary between the gas (air) and the liquid (water) is a surface. The liquid flow is assumed to be incompressible, while the gas is represented by a medium, above the liquid, in which the pressure is constant (the atmospheric pressure in general). Such a description is known to be valid for calculating the propagation in the open sea of waves with moderate amplitude, which do not break. Clearly it is not satisfactory when waves either break or hit coastal structures like offshore platforms, jetties, piers, breakwaters, etc.

Our goal here is to investigate a relatively simple two-fluid model that can handle breaking waves. It belongs to the family of averaged models, in the sense that even though the two fluids under consideration are not miscible, there exists a length scale $\epsilon$ such that each averaging volume (of size $\epsilon^3$)
contains representative samples of each of the fluids. Once the averaging process is performed, it is assumed that the two fluids share, locally, the same pressure, temperature and velocity. Such models are called homogeneous models in the literature. They can be seen as limiting cases of more general two-fluid models where the fluids can have different temperatures and velocities \cite{Ishii1975}. Let us explain why it can be assumed here that both fluids share the same temperatures and velocities. There are relaxation mechanisms that indeed tend to locally equalize these two quantities. Concerning temperatures, these are diffusion processes and provided no phenomenon is about to produce very strong gradients of temperature between the two fluids like {\it e.g.} a nuclear reaction in one of the two fluids, one can assume that the time scale on which diffusion acts is much smaller than the time scale on which the flow is averaged. Similarly, concerning the velocities, drag forces tend to locally equalize the two velocities. Define a time scale built on the mean convection velocity and a typical length scale. For flows in which the mean convection velocity is moderate, this time scale based on convection is much larger than the time scale on which velocities are equalized through turbulent drag forces. Hence, in the present model, the partial differential equations, which express conservation of mass ($1$ per fluid), balance of momentum and total energy, read as follows:
\begin{eqnarray}\label{eq:massphys}
(\alpha^+\rho^+)_t  + \div(\alpha^+\rho^+\u) &=& 0, \\ \label{eq:massphys2}
  (\alpha^-\rho^-)_t  + \div(\alpha^-\rho^-\u) &=& 0, \\ \label{eq:momentumphys}
  (\rho\u)_t + \div\left(\rho\u\otimes\u + p\I\right) &=& \rho\g, \\
  \left(\rho E\right)_t + \div\left(\rho H\u\right) &=& \rho\g\cdot\u, \label{eq:energyphys}
\end{eqnarray}
where the superscripts $\pm$ are used to denote liquid and gas respectively. Hence $\alpha^+$ and $\alpha^-$ denote the volume fraction of liquid and gas, respectively, and satisfy the condition $\alpha^+ + \alpha^-=1$. We denote by $\rho^\pm$, $\u$, $p$, $e$ respectively the density of each phase, the velocity, the pressure, the specific internal  energy, $\g$ is the acceleration due to gravity (in two space dimensions, $\g$ is equal to $(0,-g)$),  $\rho := \alpha^+\rho^+ + \alpha^-\rho^-$ is the total density, $E = e + \frac12|\u|^2$ is the specific total energy, $H := E + {p}/{\rho}$ is the specific total enthalpy. In order to close the system, we assume that the pressure $p$ is given as a function of three parameters, namely $\alpha\equiv\alpha^+-\alpha^-$, $\rho$ and $e$:
\begin{equation}\label{EOS123}
p=\mathcal{P}(\alpha, \rho, e)\,.
\end{equation}
We shall discuss in Section \ref{modelstudy} how such a function $\mathcal{P}$ is determined once the two independent equations of state $p=\mathcal{P}^\pm(\rho^\pm, e^\pm)$ are known. Equations (\ref{eq:massphys})--(\ref{EOS123}) form a closed system that we shall use to simulate aerated flows.

The main purpose of this paper is to promote a general
point of view, which may be useful for various applications dealing with violent aerated flows in ocean, offshore, coastal and arctic engineering. We do not consider here underwater explosions, where the word violent has a different meaning. The detonation of an explosive charge underwater results in an initial high-velocity shockwave through the water, in movement or displacement of the water itself and in the formation of a high-pressure bubble of high-temperature gas. This bubble expands rapidly until it either vents to the surface or until its internal pressure is exceeded by that of the water surrounding it \cite{Mehaute1995}. What we do is to follow the approach first used,
we believe, by the late Howell Peregrine and his collaborators \cite{Bredmose1,Bredmose2,Bredmose3}. The influence of the presence of 
air in wave impacts is a difficult topic. While it is usually thought that the
presence of air softens the impact pressures, recent results show that the cushioning
effect due to aeration via the increased compressibility of the air-water mixture is not necessarily a dominant
effect \cite{Bullock2007}. First of all, air may become trapped or entrained in the water in different ways, for example as
a single bubble trapped against a wall, or as a column or cloud of small bubbles. In addition, it is not clear
which quantity is the most appropriate to measure impacts. For example some researchers
pay more attention to the pressure impulse than to pressure peaks. The pressure impulse is defined as the
integral of pressure over the short duration of impact. A long time ago, Bagnold \cite{Bagnold1939} noticed that the maximum pressure and impact duration differed from one identical wave impact to the next, even in carefully controlled laboratory
experiments, while the pressure impulse appears to be more repeatable. For sure, the simple one-fluid models which are commonly
used for examining the peak impacts are no longer appropriate in the presence of air. There are few studies dealing
with two-fluid models. An exception is the work by Peregrine and his collaborators. Wood et al. \cite{Wood2000}
used the pressure impulse approach to model a trapped air pocket. Peregrine \& Thais \cite{Peregrine1996} examined the effect
of entrained air on a particular kind of violent water wave impact by considering a filling flow. Bullock et al. \cite{Bullock2001}
found pressure reductions when comparing wave impact between fresh and salt water, due to the different properties
of the bubbles in the two fluids. Indeed the aeration levels are much higher in salt water than in fresh water.
Bredmose \cite{Bredmose2005}
recently performed numerical experiments on a two-fluid system which has similarities with the one we will use below.

The novelty of the present paper is not the finite volume method used below but
rather the modelling of two-fluid flows. Since the model described below does not involve the tracking nor the
capture of a free surface, its integration is cheap from the
computational point of view. We have chosen to report here on the case of inviscid flow. Should the
viscosity effects become important, they can be taken into
account via {\it e.g.} a fractional step method. In fact, when viscous effects are important, the flow is
easier to capture from the numerical point of view.

The paper is organized as follows. Section \ref{modelstudy} provides an analytical study of the model. Section \ref{modelnum} deals with numerical simulations based on this model via a finite volume method. Two examples are shown: the dam break
problem and the drop test problem. Finally a conclusion ends the paper.

\section{Analytical study of the model}
\label{modelstudy}

\subsection{The extended equation of state}

It is shown in this section how to determine the function $\mathcal{P}(\alpha,\rho,e)$ in Eq. (\ref{EOS123}) once the two equations of state $p=\mathcal{P}^\pm(\rho^\pm, e^\pm)$ are known.
We call Eq. (\ref{EOS123}) an extended EOS, since $\mathcal{P}(-1, \rho, e)=\mathcal{P}^-(\rho, e)$ and $\mathcal{P}(1, \rho, e)=\mathcal{P}^+(\rho, e)$, where
\begin{equation}\label{EOS567}
p^\pm=\mathcal{P}^\pm(\rho^\pm, e^\pm)\,,\quad T^\pm=\mathcal{T}^\pm(\rho^\pm, e^\pm)\,,
\end{equation}
are the EOS of each fluid, with $T^\pm$ the temperature of each phase. Although our approach is totally general, we will use the following prototypical example in this paper. Assume that the
fluid denoted by the superscript $-$ is an ideal gas:
\begin{equation}\label{eq:light}
  p^- = (\gamma^- - 1) \rho^- e^-, \qquad e^- = C_V^- T^-,
\end{equation}
while the fluid denoted by the superscript $+$ obeys the stiffened gas law \cite{Cole1948,Godunov1979}:
\begin{equation}\label{eq:heavy}
  p^+ + \pi^+ = (\gamma^+ - 1) \rho^+ e^+, \qquad e^+ = C_V^+T^+ + \frac{\pi^+}{\gamma^+ \rho^+},
\end{equation}
where $\gamma^\pm$, $C_V^\pm$, and $\pi^+$ are constants. For example, pure water is well described in the vicinity of the normal conditions by taking $\gamma^+ = 7$ and $\pi^+ = 2.1\times10^9$ Pa.

Let us now return to the general case. In order to find the function $\mathcal{P}$, there are three given quantities: $\alpha\in[-1,1]$\,, $\rho>0$ and $e>0\,$. Then one solves for the four unknowns $\rho^\pm\,, e^\pm$ the following system of four nonlinear equations:
\begin{eqnarray}\label{nonlin1}
  (1+\alpha)\rho^++(1-\alpha)\rho^- &=& 2\rho\,, \\\label{nonlin2}
  (1+\alpha)\rho^+e^++(1-\alpha)\rho^-e^- &=& 2\rho\,e\,, \\\label{nonlin3}
  \mathcal{P}^+(\rho^+, e^+)-\mathcal{P}^-(\rho^-, e^-) &=& 0\,, \\\label{nonlin4}
 \mathcal{T}^+(\rho^+, e^+)-\mathcal{T}^-(\rho^-, e^-) &=& 0\,.
\end{eqnarray}
For given values of the pressure $p>0$ and the temperature $T>0$, we denote by $\mathcal{R}^\pm(p,T)$ and $\mathcal{E}^\pm(p,T)$ the solutions $(\rho^\pm, e^\pm)$ to:
\begin{equation}\label{inversion1}
\mathcal{P}^\pm(\rho^\pm, e^\pm)=p\,,\quad \mathcal{T}^\pm(\rho^\pm, e^\pm)=T\,,
\end{equation}
and then:
\begin{eqnarray}\label{inversion2}
\rho&=&\frac{1+\alpha}{2}\mathcal{R}^+(p,T)+\frac{1-\alpha}{2}\mathcal{R}^-(p,T)\,,\\
\label{inversion3}
\rho\,e&=&\frac{1+\alpha}{2}\mathcal{R}^+(p,T)\,\mathcal{E}^+(p,T)+\frac{1-\alpha}{2}\mathcal{R}^-(p,T)\,\mathcal{E}^-(p,T)\,.
\end{eqnarray}
Finally the inversion of this system of equations leads to $p=\mathcal{P}(\alpha,\rho,e)$ and $T=\mathcal{T}(\alpha,\rho,e)$.

\begin{remark}
The system (\ref{eq:massphys})--(\ref{eq:energyphys}), (\ref{eq:light})-(\ref{eq:heavy}) and (\ref{inversion1}) is a differential and algebraic equation, while the system (\ref{eq:massphys})--(\ref{EOS123}) is a partial differential equation as it is the case for a system of single fluid equations. 
\end{remark}

Concerning the prototypical case, the following generalization of (\ref{eq:light}) is considered:
\begin{equation}\label{eq:light1}
  p^- + \pi^-= (\gamma^- - 1) \rho^- e^-, \qquad e^- = C_V^- T^-+ \frac{\pi^-}{\gamma^- \rho^-}\,.
\end{equation}
This generalization, which has the additional parameter $\pi^-$, allows one to set the speed of sound to a certain value independently of $\gamma^-$, $p^-$ and $\rho^-$. 
Using computer algebra to invert (\ref{inversion2}) and (\ref{inversion3}) leads to the following expressions:
\begin{eqnarray}\label{EOSpression}
\mathcal{P}(\alpha, \rho, e) & = & (\gamma(\alpha)-1)\rho\,e-\pi(\alpha)\,, \\
\label{EOStemperature}
\mathcal{T}(\alpha, \rho, e) & = & \frac{\rho\,e-(\lambda^+(\alpha)\pi^++\lambda^-(\alpha)\pi^-)}{\rho \, C_V(\alpha)}\,,
\end{eqnarray}
where the five functions $\gamma(\alpha)$, $\pi(\alpha)$, $C_V(\alpha)$ and $\lambda^\pm(\alpha)$
are defined by
\begin{eqnarray}\label{gamma}
\frac{2}{\gamma(\alpha)-1} & = & \frac{1+\alpha}{\gamma^+-1}+\frac{1-\alpha}{\gamma^--1}\,, \\
\label{pi}
\frac{2\,\pi(\alpha)}{\gamma(\alpha)-1} & = & \frac{1+\alpha}{\gamma^+-1}\pi^++\frac{1-\alpha}{\gamma^--1}\pi^-\,, \\
\label{defCV}
\left(\frac{1+\alpha}{C_V^+(\gamma^+-1)}+\frac{1-\alpha}{C_V^-(\gamma^--1)}\right)C_V(\alpha) & = &
\frac{1+\alpha}{\gamma^+-1}+\frac{1-\alpha}{\gamma^--1}\,, \\
\label{deflambda}
\lambda^\pm(\alpha) & \equiv & \frac{1\pm\alpha}{2(\gamma^\pm-1)}
\left(1-\frac{C_V(\alpha)}{\gamma^\pm C_V^\pm}\right)\,.
\end{eqnarray}
One can easily check that one recovers the equations of state for each fluid in the limits $\alpha \to \pm 1$. Note that similar expressions can be found in Section 1.1 of \cite{Helluy2005} where a two-dimensional, compressible, two-fluid mathematical model was used to compute numerically wave breaking.

\subsection{A hyperbolic system of conservation laws}

In this section, we assume that the system of equations is solved in $\rit^2$, having in mind the numerical
computations performed below. However the extension to 3D is straightforward.
The system (\ref{eq:massphys})--(\ref{eq:energyphys}) can be written as
\begin{equation}\label{syst_abst}
  \pd{\w}{t} + \div\F(\w) = \S(\w)\,,
\end{equation}
where
\begin{equation}\label{eq:consvars}
  \w = (w_i)_{i=1}^{5} := (\alpha^+\rho^+, \alpha^-\rho^-,\;\;\rho u_1,\;\;\rho u_2,\;\;\rho E)\,,
\end{equation}
and, for every $\n\in\rit^2$,
\begin{equation}\label{flux1}
\F(\w)\cdot\n = (\alpha^+\rho^+\u\cdot \n, \alpha^-\rho^-\u\cdot \n, \rho \u\cdot \n u_1+ p n_1, \rho \u\cdot \n u_2 + pn_2, \rho H \u\cdot \n)\,,
\end{equation}
\begin{equation}\label{source}
\S(\w)=(0,0,\rho g_1,\rho g_2,\rho\g\cdot\u)\,.
\end{equation}
The Jacobian matrix $\A(\w)\cdot\n$ is defined by
\begin{equation}\label{mat_jacob}
\A(\w)\cdot\n=\pd{(\F(\w) \cdot\n)}{\w}\,.
\end{equation}

In order to compute $\A(\w)\cdot\n$, one writes Eq. (\ref{flux1}) for $\F(\w)\cdot\n$ in terms of $\w$ and $p$:
\begin{multline}\label{eq:normalAdvflux}
  \F(\w)\cdot\n =  \Bigl(w_1\frac{w_3n_1 + w_4n_2}{w_1+w_2}, w_2\frac{w_3n_1 + w_4n_2}{w_1+w_2}, w_3\frac{w_3n_1 + w_4n_2}{w_1+w_2} + p n_1, \\ w_4\frac{w_3n_1 + w_4n_2}{w_1+w_2} + pn_2, (w_5 + p)\frac{w_3n_1 + w_4n_2}{w_1+w_2}\Bigr)\,.
\end{multline}
The Jacobian matrix (\ref{mat_jacob}) then has the following expression:
\begin{eqnarray*}
 \A(\w)\cdot\n & = & \\
 & & \hspace{-3.5cm} \begin{pmatrix}
   u_n\frac{\alpha^-\rho^-}{\rho} & -u_n\frac{\alpha^+\rho^+}{\rho} & \frac{\alpha^+\rho^+}{\rho} n_1 & \frac{\alpha^+\rho^+}{\rho} n_2 & 0 \\
   -u_n\frac{\alpha^-\rho^-}{\rho} & u_n\frac{\alpha^+\rho^+}{\rho} & \frac{\alpha^-\rho^-}{\rho} n_1 & \frac{\alpha^-\rho^-}{\rho} n_2 & 0 \\
   -u_1 u_n+\pd{p}{w_1}n_1 & -u_1 u_n+\pd{p}{w_2}n_1 & u_n + u_1 n_1 + \pd{p}{w_3}n_1 & %
   u_1 n_2 + \pd{p}{w_4} n_1 & \pd{p}{w_5} n_1 \\
   -u_2 u_n+\pd{p}{w_1}n_2 & -u_2 u_n+\pd{p}{w_2}n_2 & u_2 n_1 + \pd{p}{w_3}n_2 & %
   u_n + u_2 n_2 + \pd{p}{w_4} n_2 & \pd{p}{w_5} n_2 \\
   u_n\bigl(\pd{p}{w_1} - H\bigr) & u_n\bigl(\pd{p}{w_2} - H\bigr) & %
   u_n\pd{p}{w_3} + H n_1 & u_n\pd{p}{w_4} + H n_2 & u_n\bigl(1 + \pd{p}{w_5}\bigr) \\
 \end{pmatrix}\,,
\end{eqnarray*}
where $u_n = \u \cdot \n.$

Let us now compute the five derivatives ${\partial p}/{\partial w_i}$. A systematic way of doing it is to introduce a set of five independent physical variables and here we shall take:
\begin{equation}\label{varphys}
\varphi_1=\alpha, \quad \varphi_2=p, \quad \varphi_3=T, \quad \varphi_4=u_1, \quad \varphi_5=u_2\,.
\end{equation}
The expressions of the $w_i's$ in terms of the $\varphi_j's$ are algebraic and explicit. Hence the Jacobian matrix $\partial{w_i}/\partial{\varphi_j}$ can be easily computed. Since $\partial{\varphi_j}/\partial{w_i}$ is its inverse matrix, one finds easily with the help of a computer algebra program that
\begin{eqnarray}
  \pd{p}{w_1} = \frac{\Gamma-1}{2}(u_1^2+u_2^2)+\alpha^-\rho^-\chi^-\,, \\
  \pd{p}{w_2} = \frac{\Gamma-1}{2}(u_1^2+u_2^2)+\alpha^+\rho^+\chi^+\,,
\end{eqnarray}
\begin{equation}\label{deriv_de_p}
 \pd{p}{w_3}=-(\Gamma-1)u_1\,,\quad \pd{p}{w_4}=-(\Gamma-1)u_2\,,\quad \pd{p}{w_5}=\Gamma-1\,,
\end{equation}
where
\begin{equation}\label{chipm}
\chi^\mp=\frac{1}{\rho^\pm}\frac{(c^\mp_s)^2}{\gamma^\mp-1}-\frac{1}{\rho^\mp}\frac{(c^\pm_s)^2}{\gamma^\pm-1}\,,\quad
\chi^++\chi^-=0\,,
\end{equation}
\begin{equation}\label{cscarre}
(c^\pm_s)^2\equiv C_V^\pm\gamma^\pm(\gamma^\pm-1)T=\frac{\gamma^\pm p+\pi^\pm}{\rho^\pm}\,,
\end{equation}
\begin{equation}\label{defGamma}
\Gamma-1\equiv(\gamma(\alpha)-1)\frac{\rho c_s^2}{\gamma(\alpha)p+\pi(\alpha)}\,.
\end{equation}
In Eq. (\ref{defGamma}), we have introduced the speed of sound of the mixture $c_s$, defined by
\begin{equation}
\frac{1}{\rho c_s^2}=\frac{(1+\alpha)\gamma^+}{2\rho^+(c^+_s)^2}+\frac{(1-\alpha)\gamma^-}{2\rho^-(c^-_s)^2}-
\frac{1}{\rho a^2}\,,
\label{def1surrhoc2}
\end{equation}
with
\begin{equation}
\rho a^2\equiv \frac{(1+\alpha)\rho^+(c_s^+)^2}{2(\gamma^+-1)}+\frac{(1-\alpha)\rho^-(c_s^-)^2}{2(\gamma^--1)}\,.
\label{def1sura2}
\end{equation}
Then one can show that the Jacobian matrix $ \A(\w)\cdot\n $ has three distinct eigenvalues:
\begin{equation}
  \lambda_1 = u_n - c_s, \quad
  \lambda_{2,3,4} = u_n, \quad
  \lambda_5 = u_n + c_s,
\end{equation}
These three eigenvalues are real and there is a complete set of real valued eigenvectors. The expressions of these eigenvectors can be obtained by using a computer algebra program.
\begin{remark}
If $\pi^+=0$ and $\pi^-=0$, then $c_s^2=\frac{\gamma(\alpha)p}{\rho}$ and $a^2=\frac{c_s^2}{\gamma(\alpha)-1}$.
\end{remark}
\begin{remark}
The left hand side of (\ref{def1surrhoc2}) is positive since $\rho a^2$ is bounded from below by $\frac{(1+\alpha)\rho^+(c^+_s)^2}{2\gamma^+}+\frac{(1-\alpha)\rho^-(c^-_s)^2}{2\gamma^-}$. Thus $a^2$ is seen to play the role
of the square of the enthalpy by analogy with the monofluid case.
\end{remark}

\begin{figure}
	\centering
	\subfigure[]
		{\includegraphics[width=0.9\textwidth]{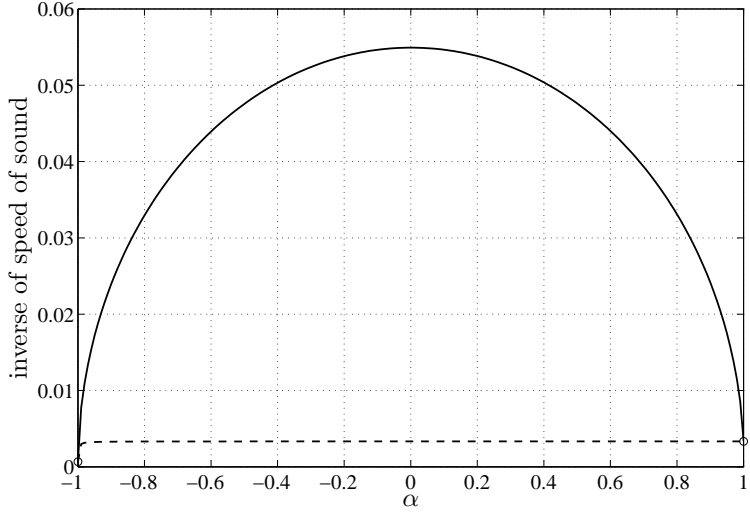}} 
		\subfigure[]
				{\includegraphics[width=0.9\textwidth]{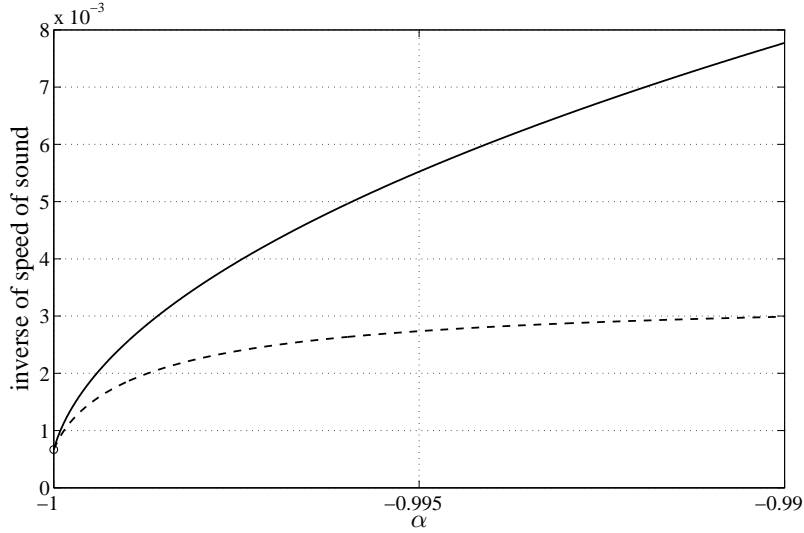}}
\caption[The speed of sound]{Speed of sound as a function of $\alpha$. In order to make the curves more visible, the inverse of the speed of sound is plotted. The end points, represented by circles, correspond to pure liquid (left) and pure gas (right). The solid line represents the inverse of the speed of sound $1/c_s$ given by equations (\ref{def1surrhoc2}) and (\ref{def1sura2}). The dashed line represents the inverse of the speed of sound $1/c_{Is}$ (\ref{cs_no_energy}) obtained without using the energy equation. (a) Full range; (b) Zoom near the pure liquid limit. The various parameters are: $\rho^+ = 1000$ kg/m$^3$, $\gamma^+ = 7$, $\pi^+ = 2.1\times10^9$ Pa,
$c^+ = 1500$ m/s and $\rho^- = 1.29$ kg/m$^3$, $\gamma^- = 1.4$, $\pi^- = 0$ Pa, $c^- = 300$ m/s.
}
	\label{fig:speed_sound}
\end{figure}

A plot of $1/c_s(\alpha)$ is given in Fig. \ref{fig:speed_sound} (see the solid line). A remarkable property is that the speed of sound exhibits a minimum. If the energy equation was not taken into consideration, this minimum would not be present (see the dashed line in Fig. \ref{fig:speed_sound}). 
Indeed, the expression for the speed of sound for homogeneous two-fluid models with the additional assumption that the flow is isentropic is 
\begin{equation}
c_{Is} = \sqrt \frac{(\alpha^-\rho^+ + \alpha^+\rho^-)(c_s^+)^2(c_s^-)^2}{\alpha^+\rho^-(c_s^-)^2 + \alpha^-\rho^+(c_s^+)^2}.
\label{cs_no_energy}
\end{equation}
This expression can be found in Appendix C of \cite{DDG2008} (see equations (C.20) and (C.21)).

\subsection{Evolution equations for the physical variables}

The system of conservation laws (\ref{eq:massphys})--(\ref{eq:energyphys}) can be transformed into a set of evolution equations for the physical variables. Let us introduce the entropy function $s(\x,t)$ defined by (compare with Eq. (\ref{nonlin2}))
$$ 2\rho\,s = (1+\alpha)\rho^+s^++(1-\alpha)\rho^-s^-. $$
\begin{proposition}
Continuous solutions to (\ref{eq:massphys})--(\ref{eq:energyphys}) satisfy
\begin{eqnarray}\label{eq:vitesse}
\u_t + \u\cdot\nabla\u + \frac{1}{\rho}\nabla p &=& \g\,, \\\label{eq:pression}
  p_t + \u\cdot\nabla p + \rho c_s^2\div \u &=& 0\,, \\\label{eq:alpha}
  \alpha_t + \u\cdot\nabla \alpha + (1-\alpha^2)\,\delta\,\div \u &=& 0\,, \\
  s_t + \u\cdot\nabla s &=& 0\,, \label{eq:entropie}
  \end{eqnarray}
  where $c_s^2$ is given by (\ref{def1surrhoc2})-(\ref{def1sura2}) and $\delta$ is given by
\begin{equation}\label{def_delta}
\delta\equiv \frac{1}{2}\frac{\rho c_s^2(\gamma^-\pi^+-\gamma^+\pi^-)}{\rho^+\rho^-(c_s^+)^2(c_s^-)^2}\,.
\end{equation}
\end{proposition}

\begin{remark}
For pure fluids ($\alpha=\pm 1$), Eq. (\ref{eq:alpha}) is no longer relevant and $\delta$ is not needed. One can check
that the speed of sound $c_s$ is then equal to the expected speed of sound ($c_s^+$ or $c_s^-$) for pure fluids.
\end{remark}

The balance of entropy (\ref{eq:entropie}) comes from the balance
\begin{equation}\label{eq:entropy}
 (\rho s)_t  + \div (\rho s \u) = 0.
\end{equation}
Adding together Eqs (\ref{eq:massphys}) and (\ref{eq:massphys2}) leads to
\begin{equation}\label{eq:total_mass}
\rho_t + \div (\rho \u) =  0.
\end{equation}
Combining Eqs (\ref{eq:entropy}) and (\ref{eq:total_mass}) leads to Eq. (\ref{eq:entropie}).

\begin{remark}
Subtracting Eq. (\ref{eq:massphys}) from Eq. (\ref{eq:massphys2}) leads to
\begin{equation}\label{chi}
(\rho\chi)_t + \div (\rho\chi \u) =  0\,, \quad \mbox{with} \;\; \chi = \frac{\alpha^+\rho^+-\alpha^-\rho^-}
{\rho}\,.
\end{equation}
In the case of smooth solutions, we obtain that
$$ \chi_t + \u\cdot\nabla\chi = 0 \,, $$
which is an alternative to Eq. (\ref{eq:alpha}).
\end{remark}


\subsection{Pure fluid limit}
\label{sec:prop}

The two-fluid model described in the present paper is based on the volume fraction of liquid and gas. In some situations, this volume fraction can have sharp gradients. Consider for example a tanh-type distribution of $\alpha$ along the vertical axis with essentially pure gas at the top, pure liquid at the bottom and a middle layer where $\alpha$ goes rapidly from $-1$ to 1. One can even consider the limiting case where the transition is discontinuous. In this section we study this limit and we show that the two-fluid model degenerates into the classical water-wave equations. In other words one has an interface separating two pure fluids. So the well-known water-wave equations are a by-product of the two-fluid system under investigation. A similar type of limit in the case of a continuously stratified incompressible fluid degenerating into a two-layer incompressible fluid was considered by James \cite{James2001}.

In the rest of this section, it is assumed that there are no shocks. 
Consider the 3D case where $\alpha$ is either $1$ or $-1$. More precisely let
\begin{equation}\label{alpha_i}
    \alpha := 1 - 2{\mathcal H}(z-\eta(\x,t))\,, \quad \x = (x_1,x_2)\,,
\end{equation}
where ${\mathcal H}$ is the Heaviside step function, $z$ the vertical coordinate and $x_1,x_2$ the horizontal coordinates. Physically this substitution means that we consider two pure fluids separated by an interface. It follows that
$$
  \alpha^+ \alpha^- = 0\,, \quad 1-\alpha^2 = 0\,.
$$

Substituting the expression (\ref{alpha_i}) into the equation (\ref{eq:alpha}) gives
\begin{equation*}
    \eta_t + \u_h\cdot\nabla_h\eta = w\,,
\end{equation*}
where $\u_h = (u_1,u_2)$, $\nabla_h = (\partial_{x_1}, \partial_{x_2})$ and $w$ is the vertical velocity.

This equation simply states that there is no mass flux across the interface. Incidentally this is no longer true in the case of shock waves.  Integrating the conservation of momentum equation (\ref{eq:momentumphys}) inside a volume moving with the
flow and enclosing the interface, and using the fact that there is no mass flux across the interface simply leads to the fact that there is no pressure jump across the interface. In other words, the pressure is continuous across the interface. Integrating the entropy equation
inside the same volume enclosing the interface and using the fact there is no mass flux across the interface does not lead
to any new information.

One can now write Eqs (\ref{eq:massphys2})--(\ref{eq:energyphys}) in each fluid by taking $\alpha^\pm = 1$, either in the conservative form
\begin{eqnarray}
(\rho^\pm)_t + \div(\rho^\pm \u^\pm) &=& 0\,, \\
(\rho^\pm\u^\pm)_t + \div(\rho^\pm \u^\pm \otimes \u^\pm)+ \nabla p^\pm &=& \rho^\pm\g\,, \\
(\rho^\pm s^\pm)_t  + \div (\rho^\pm s^\pm \u^\pm) = 0\,,
\end{eqnarray}
(see Whitham \cite{Whitham1999} for example for the last equation) or in the more classical form
\begin{eqnarray}\label{linclass1}
	\rho^\pm_t+(\u^\pm\cdot\nabla)\rho^\pm + \rho^\pm \div\u^\pm & = & 0\,, \\ \label{linclass2}
	\u^\pm_t+(\u^\pm\cdot\nabla)\u^\pm + \frac{\nabla p^\pm}{\rho^\pm} &=& \g\,, \\ \label{linclass3}
	s^\pm_t+\u^\pm\cdot\nabla s^\pm &=& 0\,.
\end{eqnarray}
In these two systems, the superscripts $+$ and $-$ are used for the heavy fluid (below the interface) and the light fluid (above the interface) respectively.

The system of equations we derived is nothing else than the system of a discontinuous two-fluid system with an interface
located at $z=\eta(\x,t)$. Along the interface, one has the kinematic and dynamic boundary conditions
\begin{eqnarray}
	\eta_t + \u_h^\pm\cdot\nabla_h\eta &=& w^\pm\,, \\
	p^- &=& p^+\,.
\end{eqnarray}

This simple computation shows an important property of our model: it automatically degenerates into a discontinuous two-fluid system where two pure compressible phases are separated by an interface. This limit has interesting consequences. In
particular, interfacial flows develop waves along the interface and these waves are usually dispersive. Therefore one
can also expect dispersive waves to exist in the two-fluid model. Since the emphasis of the present paper is the study of large-amplitude disturbances, the derivation of the dispersion relation for the two-fluid model is left for future work.
Note however that preliminary results can be found in \cite{DDG2008}. Even the question of which rest state one must consider is
not trivial. 

\section{Simulations of aerated violent flows}\label{modelnum}
\subsection{A finite-volume discretization of the model}

Here we describe the discretization of the model (\ref{eq:massphys})--(\ref{eq:energyphys}) by a standard cell-centered finite volume method. The computational domain $\Omega\subset\R^d$ is triangulated into a set of control volumes: $\Omega=\cup_{K\in\T} K$.
We start by integrating equation (\ref{syst_abst}) on $K$:
\begin{equation}\label{eq:conservlaw}
	\od{}{t}\int_{K} \w \;d\Omega + \sum_{L\in\N(K)}\int_{K\cap L}\F(\w)\cdot\n_{KL} \;d\sigma
  	  = \int_{K}\S(\w) \;d\Omega\,,
\end{equation}
where $\n_{KL}$ denotes the unit normal vector on $K\cap L$ pointing into $L$ and $
  \N(K) = \set{L\in\T: \area(K\cap L) \neq 0}\,.
$
Then, setting
\begin{equation*}
 \w_K(t) := \frac{1}{\vol(K)}\int_{K} \w(\x,t) \;d\Omega \;,
\end{equation*}
we approximate (\ref{eq:conservlaw}) by
\begin{equation}\label{a_resoudre}
	\od{\w_K}{t} + \sum_{L\in\N(K)} \frac{\area(L\cap K)}{\vol(K)} \Phi(\w_K, \w_L; \n_{KL}) =  \S(\w_K)\;,
\end{equation}
where the numerical flux $$\Phi(\w_K, \w_L; \n_{KL}) \approx\frac{1}{\area(L\cap K)}\int_{K\cap L}\F(\w)\cdot\n_{KL} \;d\sigma$$ is explicitly computed by the FVCF formula of Ghidaglia {\it et al.} \cite{Ghidaglia2001}:
\begin{equation}\label{CFFV}
\Phi(\vv, \w; n)=\frac{\F(\vv) \cdot\n+\F(\w) \cdot\n}{2}-\sgn(\A_n(\mu(\vv,\w)))\frac{\F(\w) \cdot\n-\F(\vv) \cdot\n}{2}\,.
\end{equation}
Here $\vv$ and $\w$ are dummy variables. The Jacobian matrix $\A_n(\mu)$ is defined in (\ref{mat_jacob}), $\mu(\vv,\w)$ is an arbitrary mean between $\vv$ and $\w$ (for example $\mu(\vv,\w)=(1/2)(\vv+\w)$) and $\sgn(M)$ is the matrix whose eigenvectors are those of $M$ but whose eigenvalues are the signs of that of $M$. As explained in \cite{Ghidaglia2001}, this method is able to model discontinuities such as 
shock waves and sharp interfaces.

So far we have not discussed the case where a control volume $K$ meets the boundary of $\Omega$. Here we shall only consider the case where this boundary is a wall and from the numerical point of view, we only need to find the normal flux $\F\cdot\n$. Since $
  \u(\x,t)\cdot\n = 0$ for $\x\in\partial\Omega\,,
$ we have
\begin{equation*}
  \left.(\F\cdot\n)\right|_{\x\in\partial\Omega} = (0, 0, p_b \n, 0), \quad p_b := \left.p\right|_{\x\in\partial\Omega}\,,
\end{equation*}
and following Ghidaglia and Pascal \cite{Ghidaglia2005}, we can take  $p_b = p + \rho u_n c_s,$
where the right-hand side is evaluated in the control volume $K$.

\begin{rem}
 In order to turn (\ref{a_resoudre}) into a numerical algorithm, we must at least perform time discretization and give an expression for $\mu(\vv,\w)$. Since this matter is standard, we do not give the details here but instead refer to Dutykh \cite{Dutykh2007a}. Let us also notice that formula (\ref{a_resoudre}) leads to a first-order scheme but in fact we use a MUSCL technique to achieve higher accuracy in space \cite{leer2006}.
\end{rem}


\subsection{Numerical results}
\label{modelres}


In order to check the accuracy of our second-order scheme on smooth solutions and its robustness against discontinuous solutions, we have performed the classical test cases for which we refer to \cite{Dutykh2007a}. The most famous test case is that of Sod's shock tube.
We report here on some of the situations which have motivated this study.
\subsubsection{Thermodynamics constants}

  The constants $C_V^\pm$ can be calculated after simple algebraic manipulations of equations (\ref{eq:light}), (\ref{eq:heavy}) and matching with experimental values at normal conditions:
  \begin{equation*}
    C_V^- \equiv \frac{p_0}{(\gamma^- -1)\rho^-_0 T_0},
  \end{equation*}
  \begin{equation*}
    C_V^+ \equiv \frac{\gamma^+ p_0 + \pi^+}{(\gamma^+-1)\gamma^+\rho_0^+ T_0}.
  \end{equation*}
  For example, for an air/water mixture under normal conditions we have the values given in Table~\ref{tab:diphaseparams}.

\begin{table}
	\begin{center}
		\begin{tabular}{cc}
  		\hline\hline
  		\textit{parameter} & \textit{value} \\
      \hline
      	$p_0$ & $10^5$ $Pa$ \\
      \hline
        $\rho^+_0$ & $10^3$ $kg/m^3$ \\
      \hline
        $\rho^-_0$ & $1.29$ $kg/m^3$ \\
      \hline
        $T_0$ & $300$ $K$ \\
      \hline
        $\gamma^-$ & $1.4$ \\
      \hline
        $\gamma^+$ & $7$ \\
      \hline
        $\pi^+$ & $2.1\times 10^9$ $Pa$ \\
      \hline
        $C_V^+$ & $166.72$ $\frac{J}{kg\cdot K}$ \\
      \hline
        $C_V^-$ & $646.0$ $\frac{J}{kg\cdot K}$ \\
      \hline\hline
		\end{tabular}
	  \caption[Parameters for an air/water mixture under normal conditions]%
	  {Values of the parameters for an air/water mixture under normal conditions.}
	  \label{tab:diphaseparams}
  \end{center}
\end{table}

The sound velocities in each phase are given by the following formulas:
\begin{equation}\label{eq:soundspeed}
  (c_s^-)^2 = \frac{\gamma^- p^-}{\rho^-}, \qquad
  (c_s^+)^2 = \frac{\gamma^+ p^+ + \pi^+}{\rho^+}.
\end{equation}

In the two test cases described below, we use a very high value for the acceleration due to gravity: $g=100$ ms$^{-2}$. The only motivation is to accelerate the dynamics. All results are
presented with physical dimensions. For example, the $1\times 1$ box used for the computations corresponds to a 1 m by 1 m box.

\subsubsection{Falling water column}

\begin{figure}[htbp]
\centering
\psfrag{A}{$\alpha^+ = 0.9$}
\psfrag{B}{$\alpha^- = 0.1$}
\psfrag{C}{$\alpha^+ = 0.1$}
\psfrag{D}{$\alpha^- = 0.9$}
\psfrag{0}{$0$}
\psfrag{0.3}{$0.3$}
\psfrag{0.65}{$0.65$}
\psfrag{0.7}{$0.7$}
\psfrag{0.05}{$0.05$}
\psfrag{1}{$1$}
\psfrag{0.9}{$0.9$}
\psfrag{g}{$\g$}
\includegraphics[width=12cm]{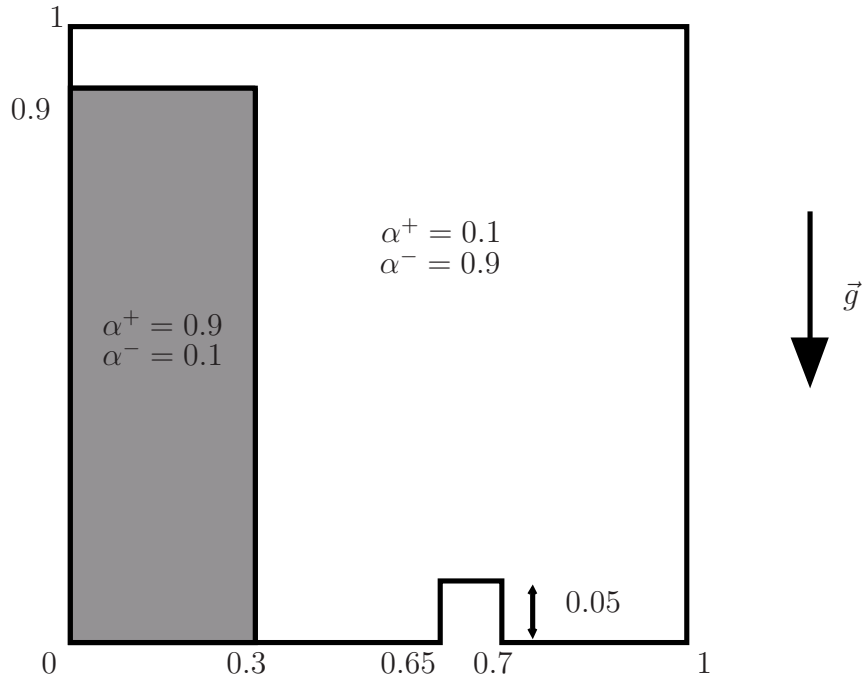}
\caption[Geometry and initial condition for falling water]{Falling water column test case. Geometry and initial condition. All the
values for $\alpha^\pm$ are at time $t=0$.}
\label{fig:falling_water}
\end{figure}

\begin{figure}
	\centering
	\subfigure[$t = 0.005$ s]%
	{\includegraphics[width=0.46\textwidth]{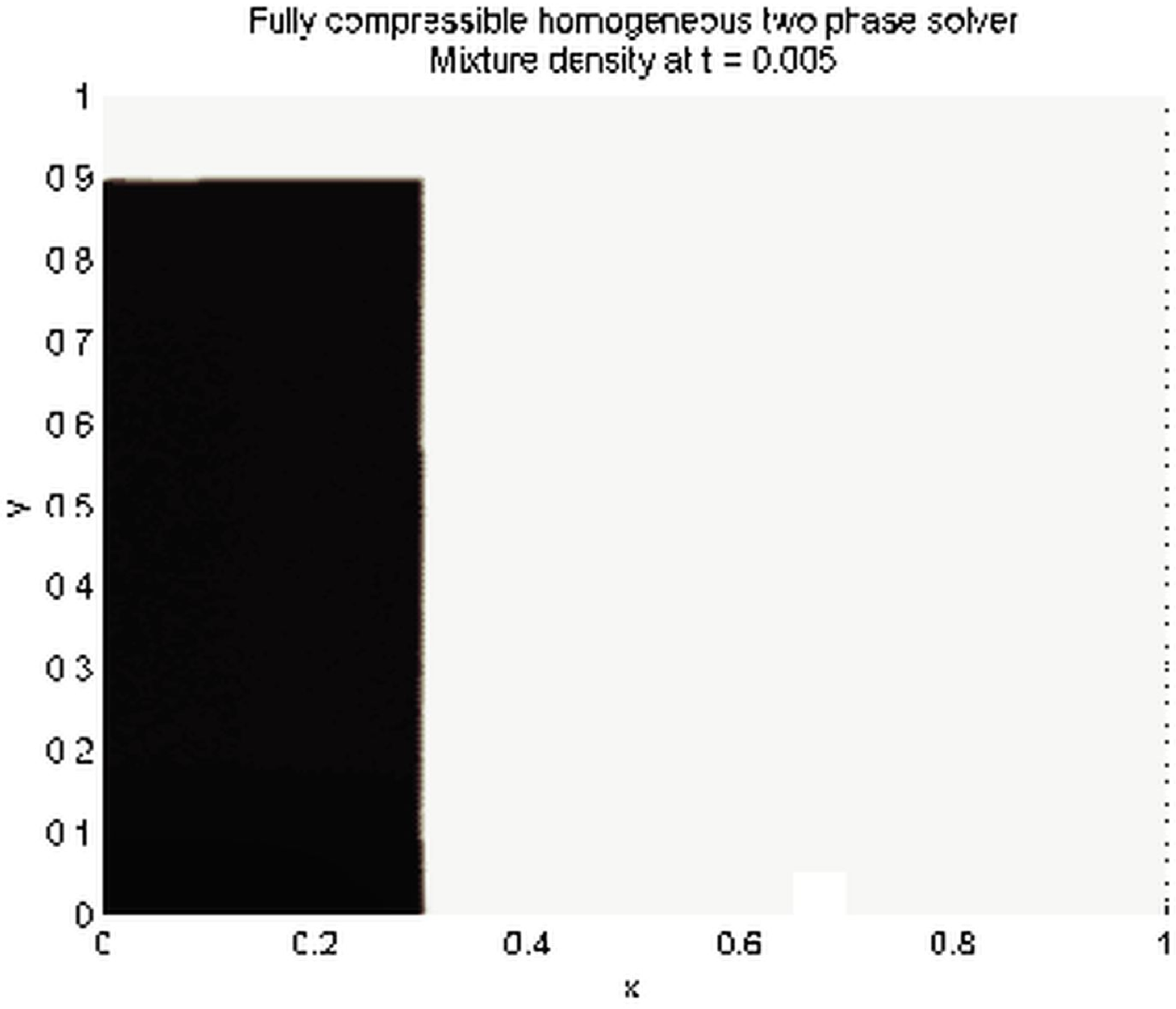}} \quad
	\subfigure[$t = 0.06$ s]%
	{\includegraphics[width=0.46\textwidth]{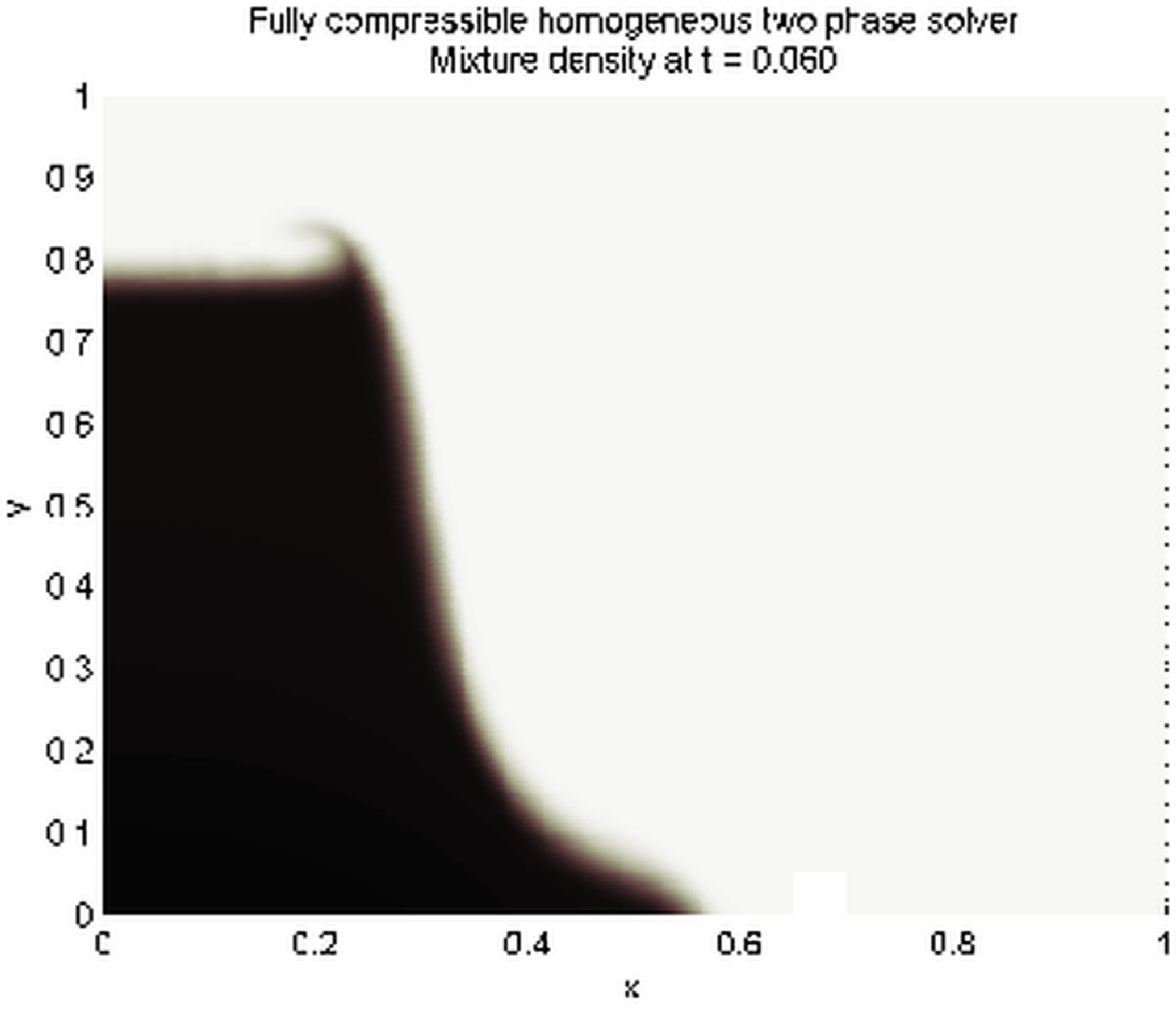}}
	\caption[The beginning of column dropping process]{Falling water column test case. Initial condition and the beginning of the column collapse.}
	\label{fig:debutSplash}
\end{figure}

\begin{figure}
	\centering
	\subfigure[$t = 0.15$ s]%
	{\includegraphics[width=0.46\textwidth]{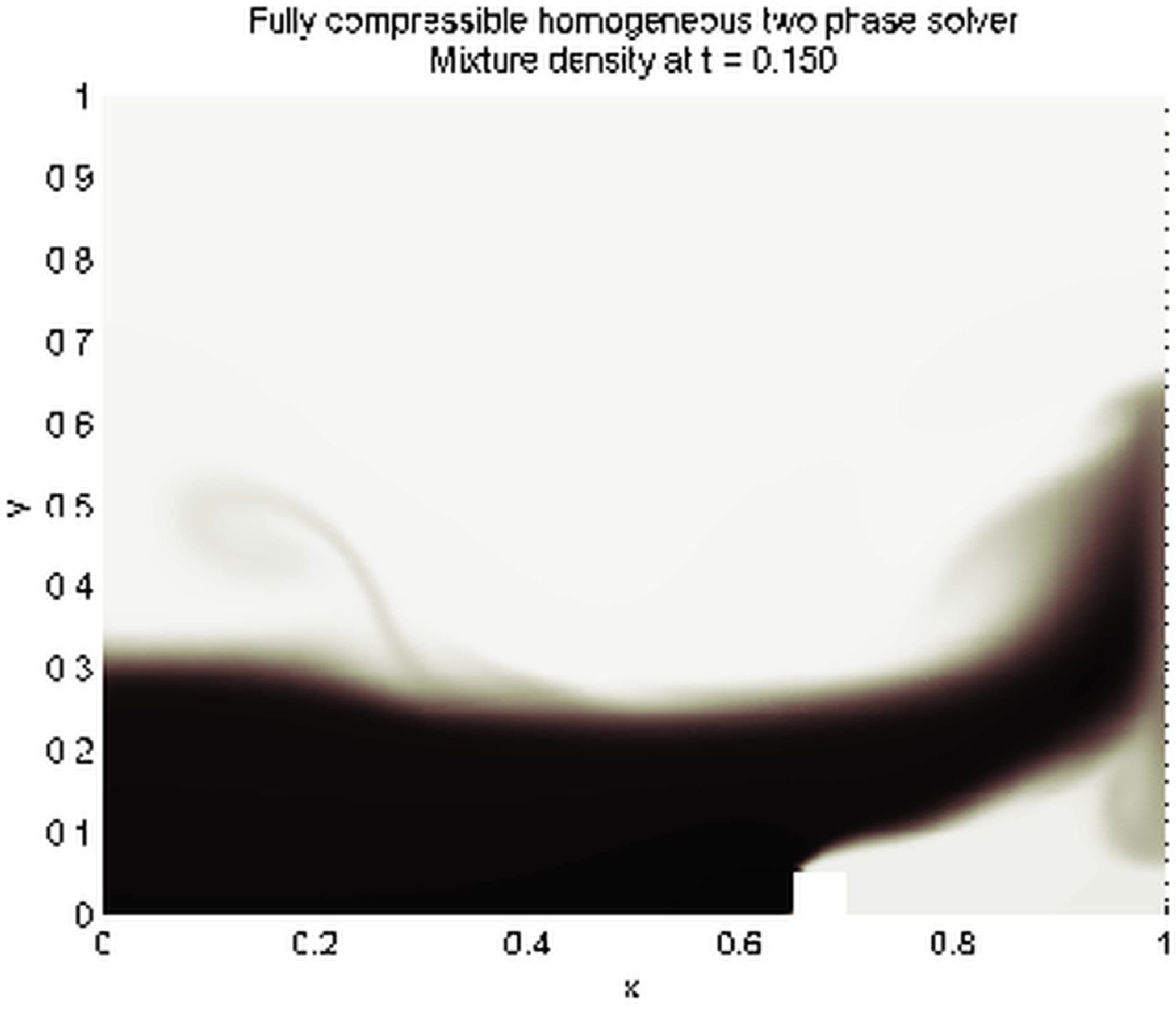}} \quad
	\subfigure[$t = 0.175$ s]%
	{\includegraphics[width=0.46\textwidth]{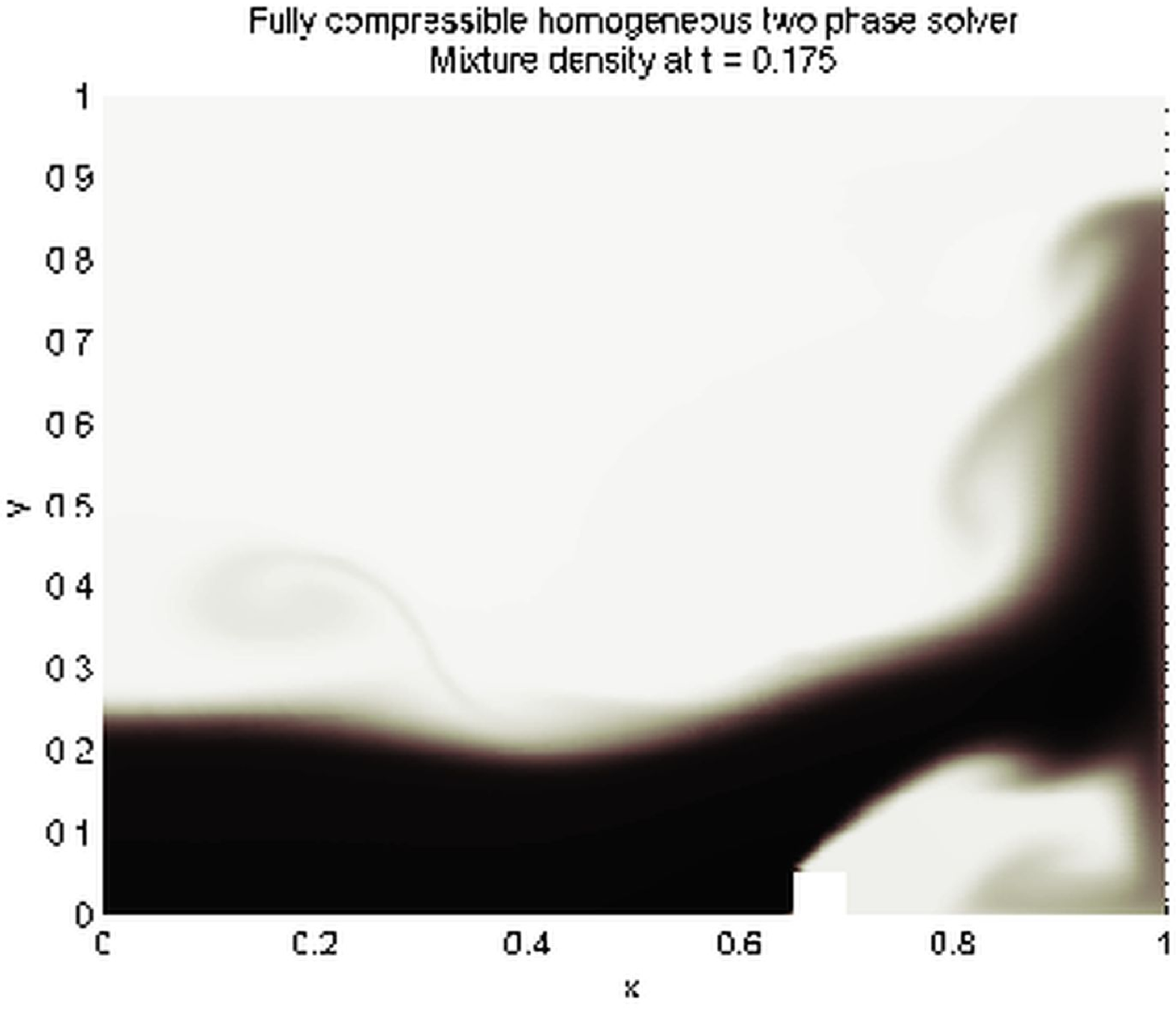}}
	\caption[Water strikes the wall - I]{Falling water column test case. Water hits the wall.}
	\label{fig:nextSplash}
\end{figure}

\begin{figure}
	\centering
	\subfigure[$t = 0.5$ s]%
	{\includegraphics[width=0.46\textwidth]{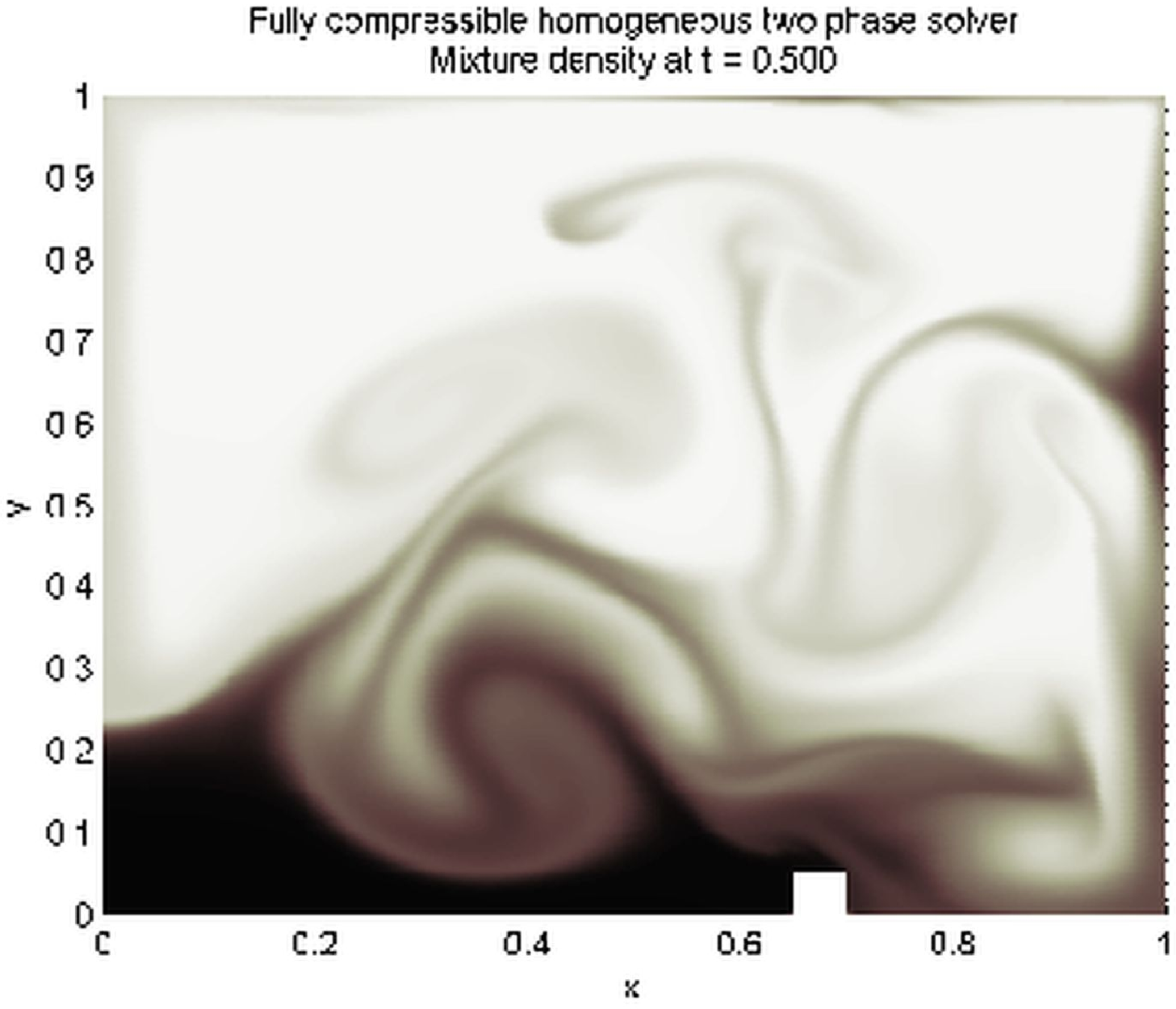}} \quad
	\subfigure[$t = 0.675$ s]%
	{\includegraphics[width=0.46\textwidth]{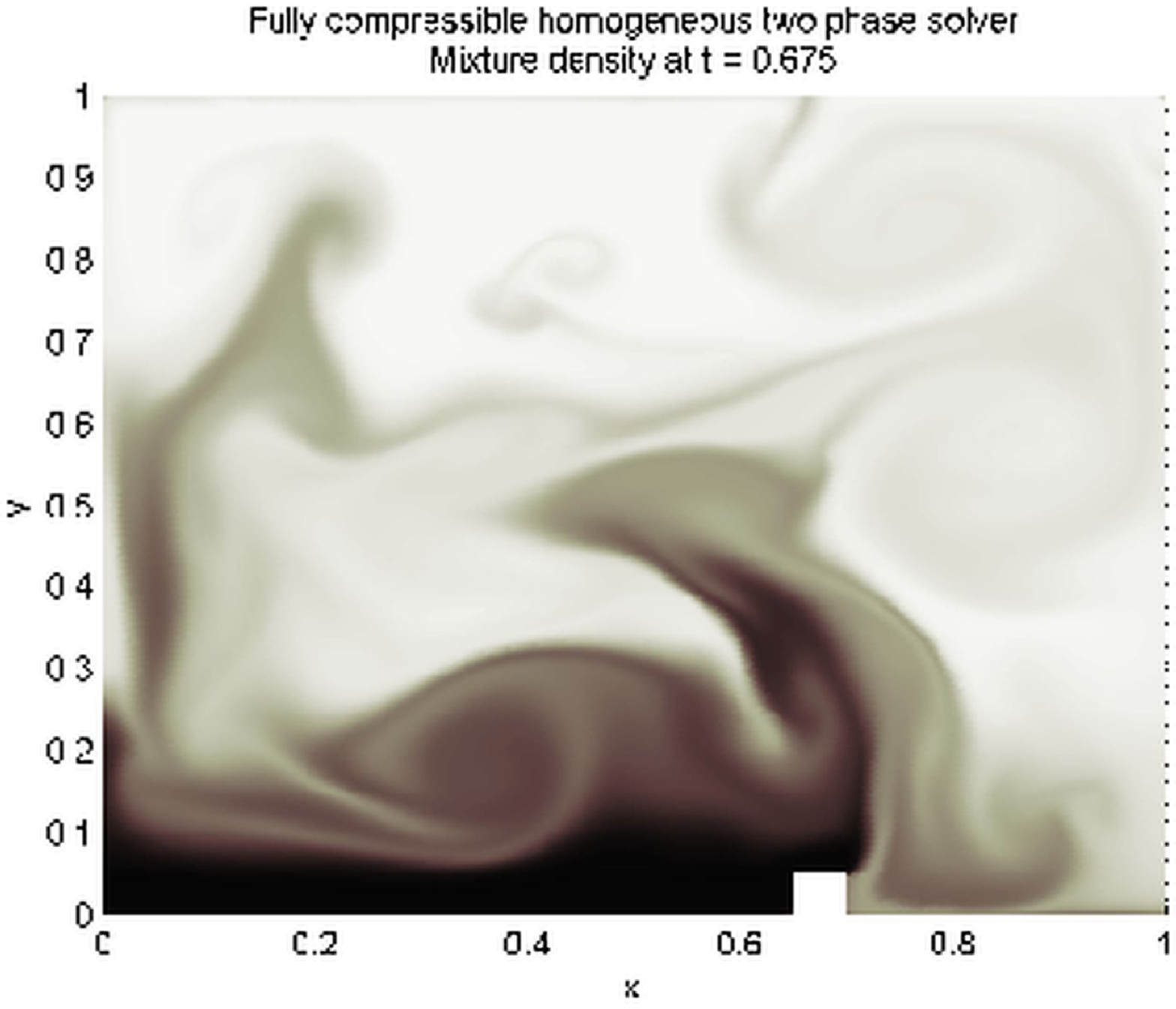}}
	\caption[Turbulent mixing process]{Falling water column test case. Turbulent mixing process.}
	\label{fig:lastSplash}
\end{figure}

\begin{figure}
	\centering
		\includegraphics[width=0.80\textwidth]{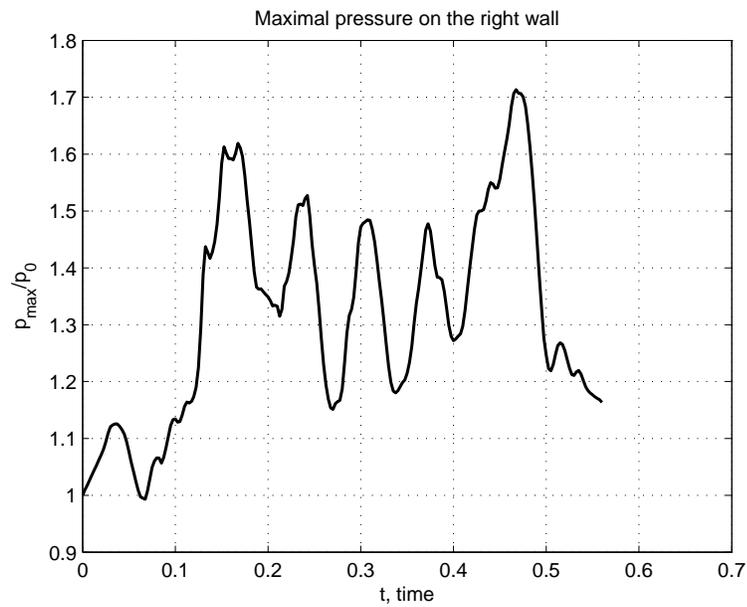}
	\caption{Maximal pressure on the right wall as a function of time. Case of a heavy gas.}
	\label{fig:wallpress1}
\end{figure}

\begin{figure}
	\centering
		\includegraphics[width=0.80\textwidth]{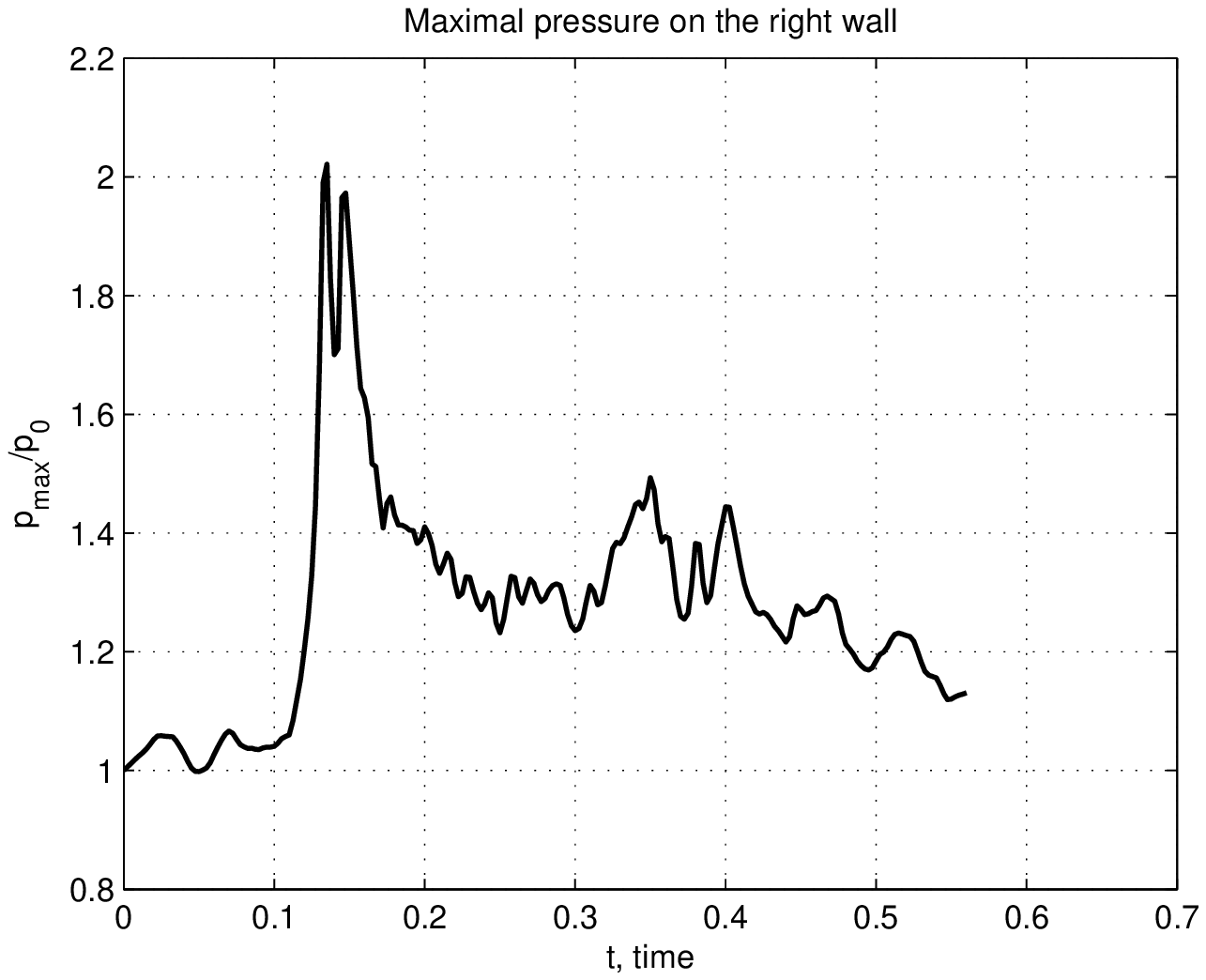}
	\caption{Maximal pressure on the right wall as a function of time. Case of a light gas.}
	\label{fig:wallpress2}
\end{figure}

The geometry and initial condition for this test case are shown on \figurename~\ref{fig:falling_water}. Initially the velocity field is taken to be zero. At time $t=0$, the volume fraction of gas is 0.9 (white area) while the volume fraction of water is 0.9 (dark area).
The values of the other parameters are given in Table \ref{tab:diphaseparams}. The mesh used in this computation contained about $108 000$ control volumes (in this case they were triangles). The results of this simulation are presented on Figures \ref{fig:debutSplash}--\ref{fig:lastSplash}.
\figurename~\ref{fig:wallpress1} shows the maximal pressure on the right wall as a function of time:
\begin{equation*}
  t \longmapsto \max_{(x,y) \in 1\times[0,1]} p(x,y,t).
\end{equation*}
We performed another computation for a mixture with $\alpha^+ = 0.05$, $\alpha^- = 0.95$. The pressure is recorded as well and plotted in \figurename~\ref{fig:wallpress2}. One can see that the peak value is higher and the impact is more localized in time.

\begin{figure}
	\centering
	\subfigure[$t = 0.1$ s]%
	{\includegraphics[width=0.46\textwidth]{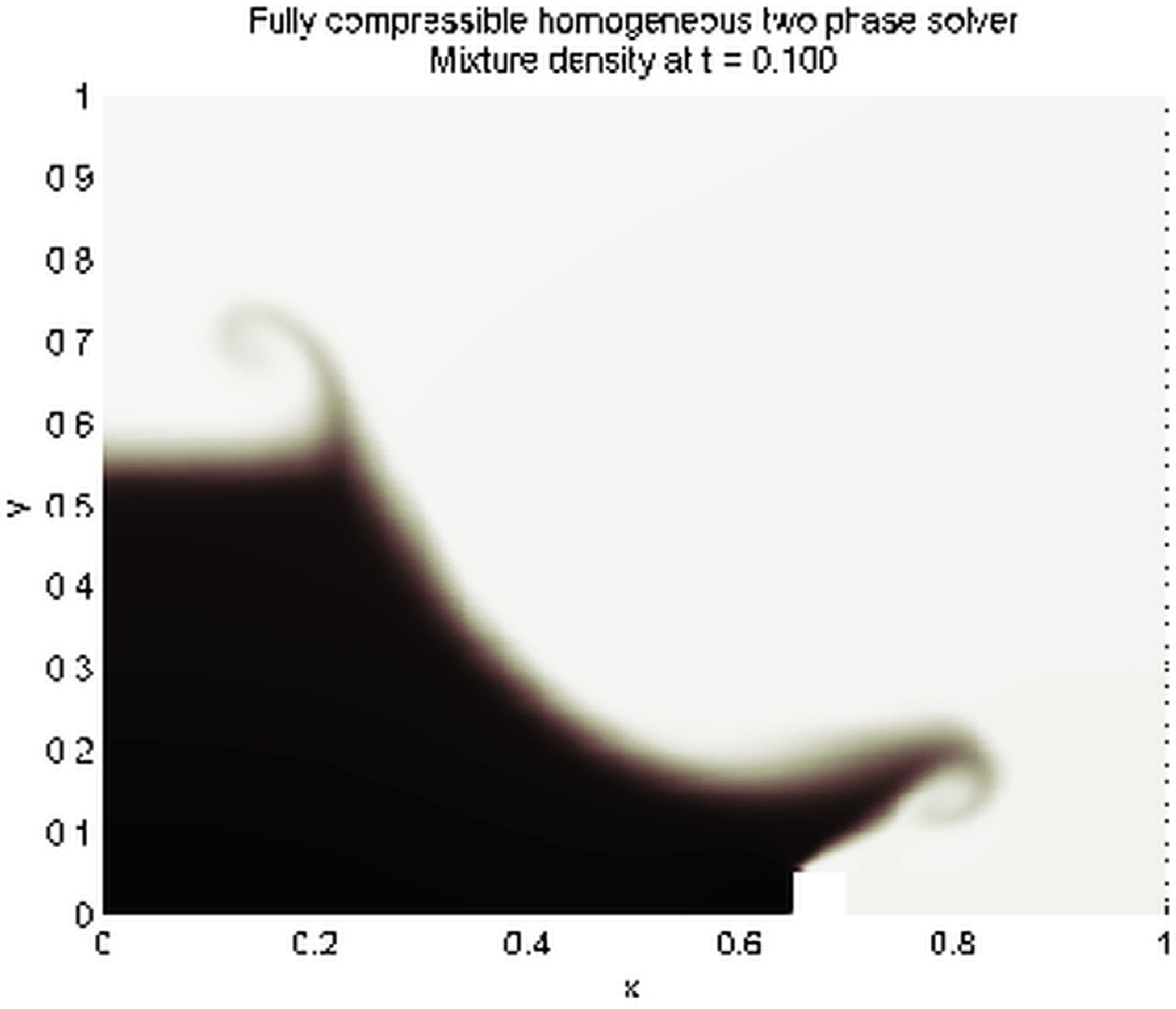}} \quad
	\subfigure[$t = 0.125$ s]%
	{\includegraphics[width=0.46\textwidth]{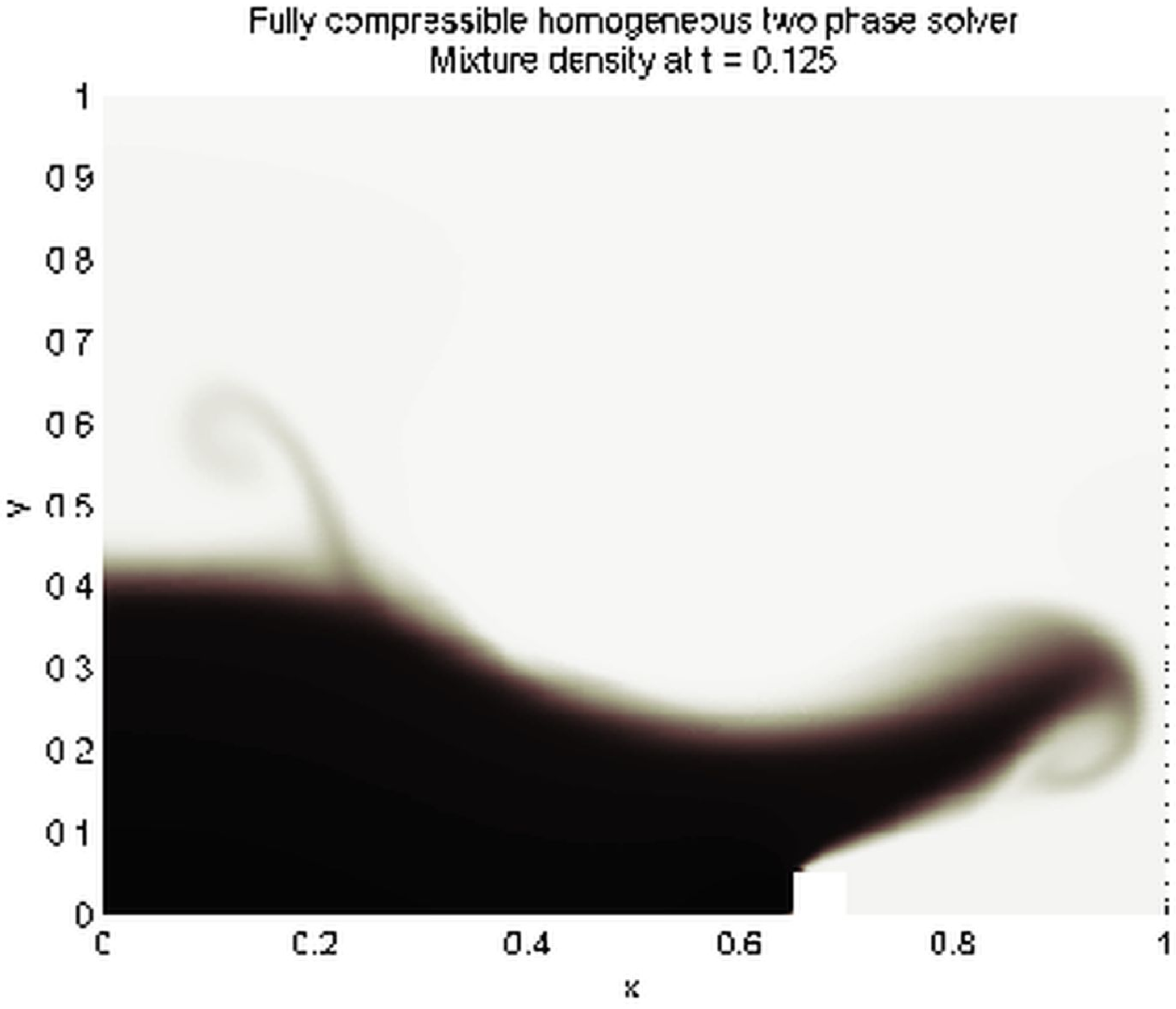}}
	\caption[Splash creation due to the interaction with step]{Falling water column test case. Splash formation due to the interaction with the step.}
\end{figure}

\begin{figure}
	\centering
	\subfigure[$t = 0.2$ s]%
	{\includegraphics[width=0.46\textwidth]{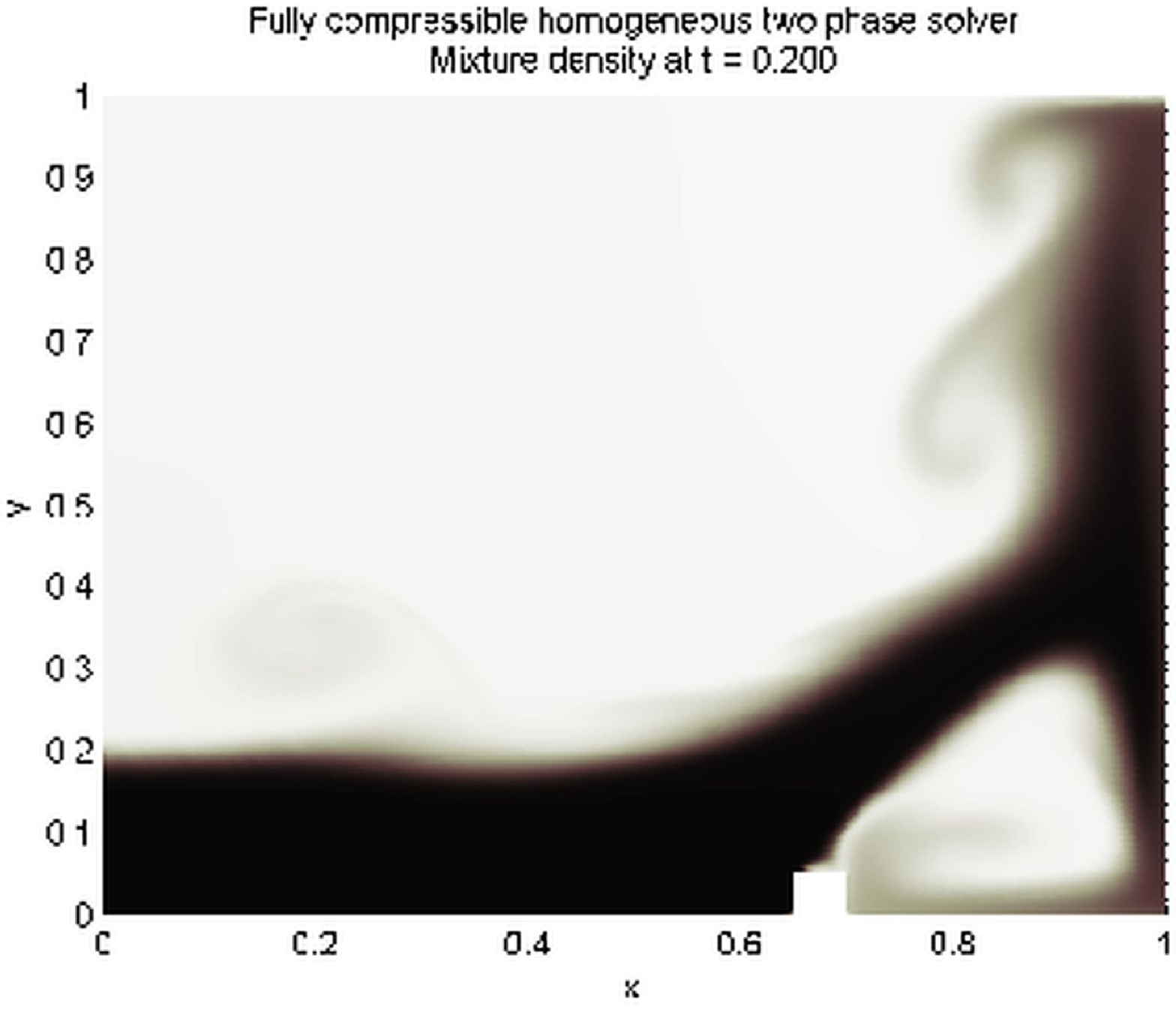}} \quad
	\subfigure[$t = 0.225$ s]%
	{\includegraphics[width=0.46\textwidth]{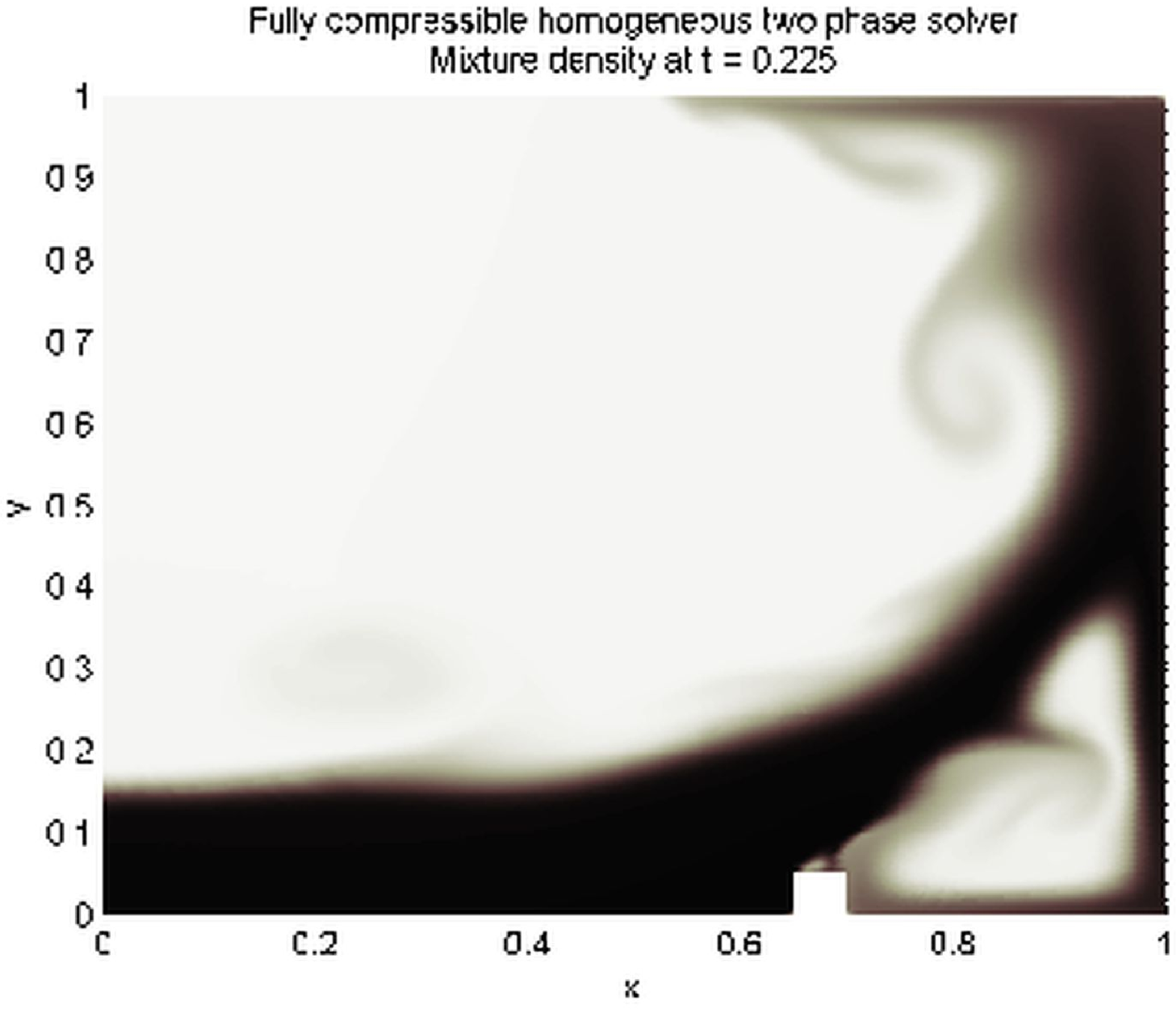}}
	\caption[Water strikes the wall - II]{Same as Fig. \ref{fig:nextSplash} at later times.}
\end{figure}

\begin{figure}
	\centering
	\subfigure[$t = 0.3$ s]%
	{\includegraphics[width=0.46\textwidth]{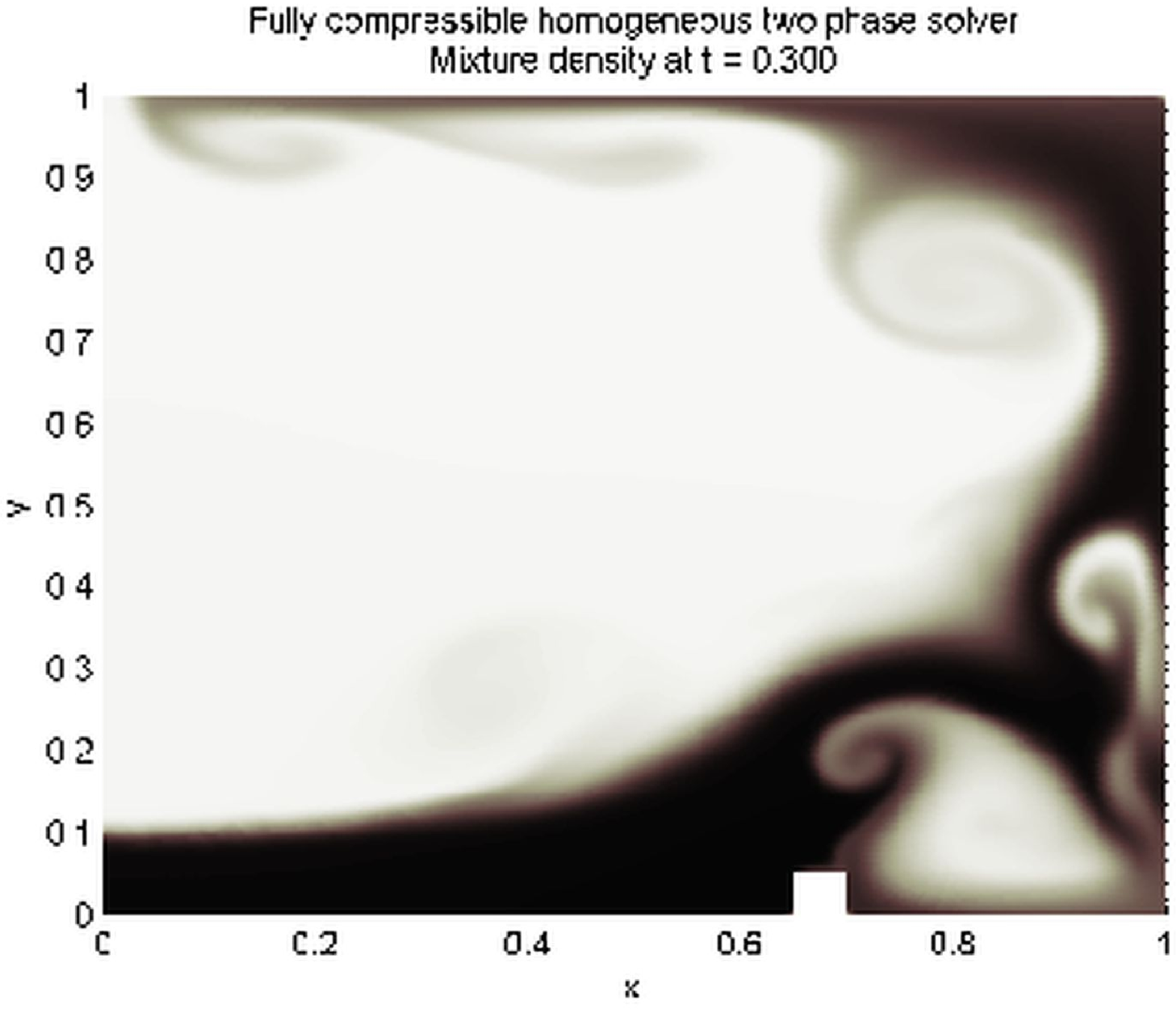}} \quad
	\subfigure[$t = 0.4$ s]%
	{\includegraphics[width=0.46\textwidth]{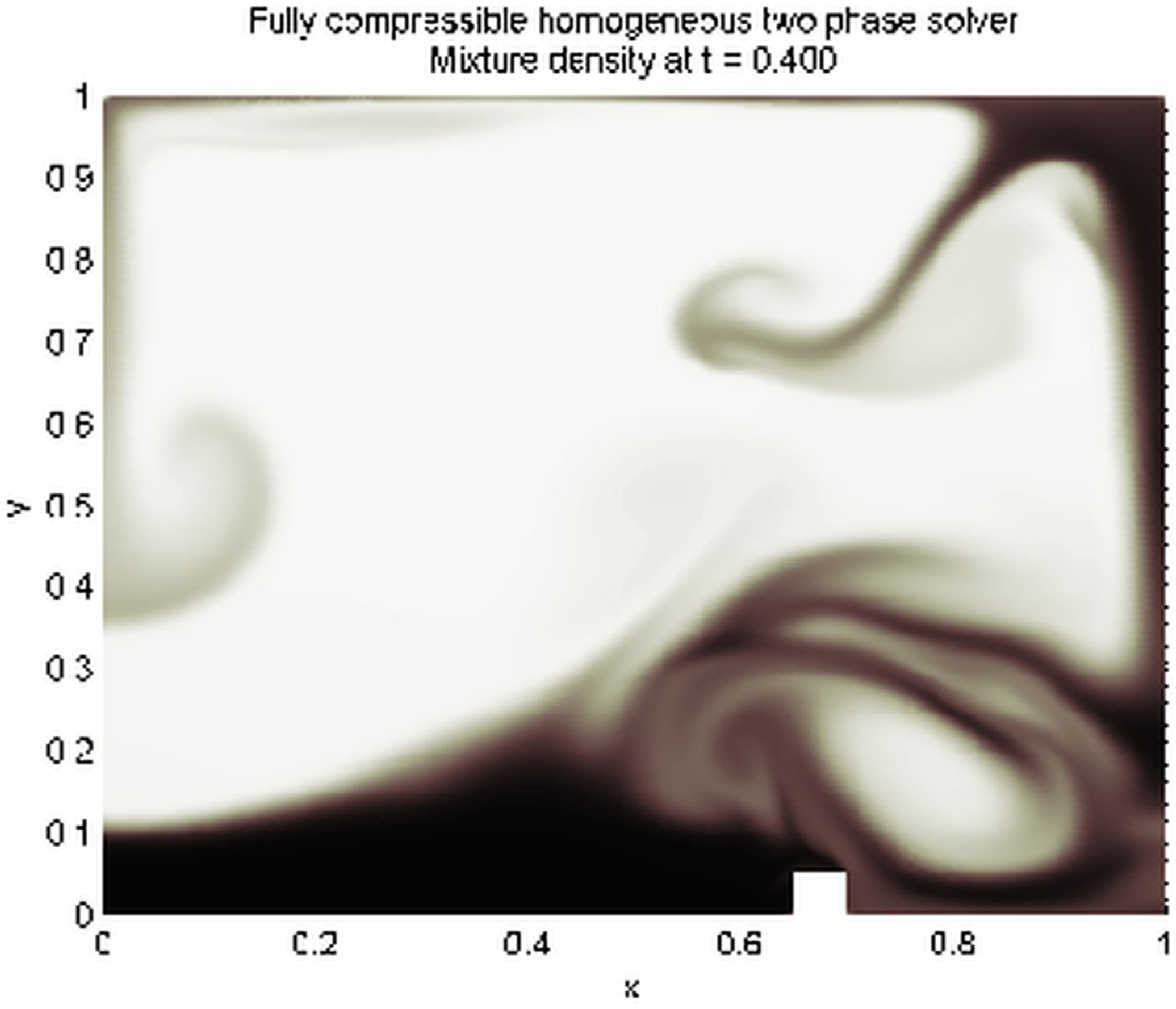}}
	\caption[Splash is climbing the wall]{Falling water column test case. The splash is climbing the wall.}
\end{figure}

\subsubsection{Water drop test case}

\begin{figure}[htbp]
\centering
\psfrag{A}{$\alpha^+ = 0.1$}
\psfrag{B}{$\alpha^- = 0.9$}
\psfrag{C}{$\alpha^+ = 0.9$}
\psfrag{D}{$\alpha^- = 0.1$}
\psfrag{0}{$0$}
\psfrag{0.5}{$0.5$}
\psfrag{0.7}{$0.7$}
\psfrag{1}{$1$}
\psfrag{g}{$\g$}
\psfrag{R}{$R = 0.15$}
\includegraphics[width=13cm]{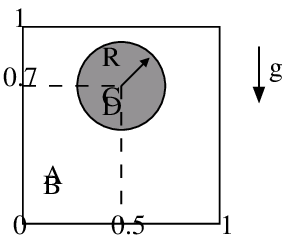}
\caption[Geometry and initial condition for water drop test case]{Geometry and initial condition for water drop test case. All the values
for $\alpha\pm$ are at time $t=0$.}
\label{fig:drop_water}
\end{figure}

\begin{figure}
	\centering
	\subfigure[$t = 0.005$ s]%
	{\includegraphics[width=0.46\textwidth]{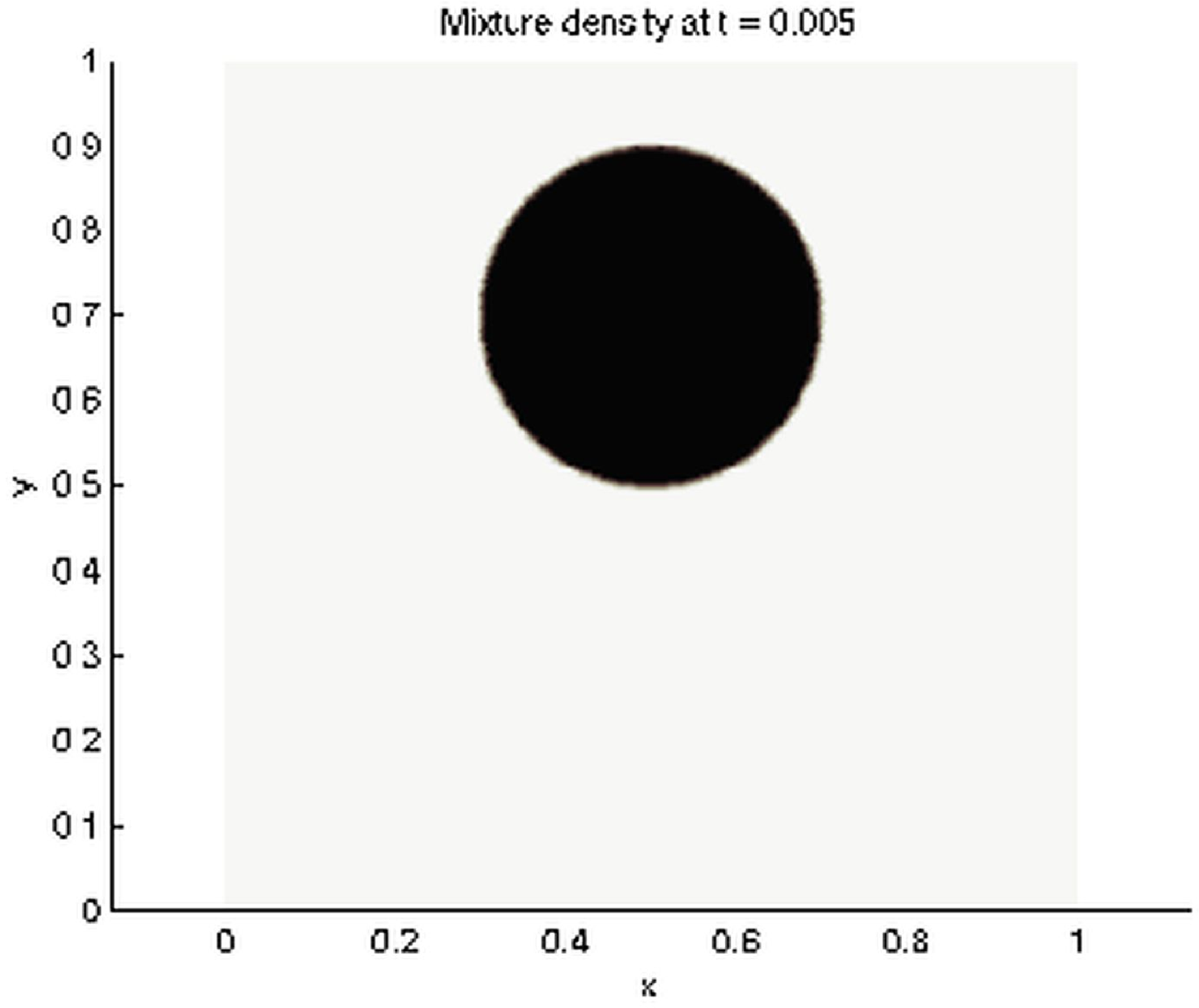}} \quad
	\subfigure[$t = 0.075$ s]%
	{\includegraphics[width=0.46\textwidth]{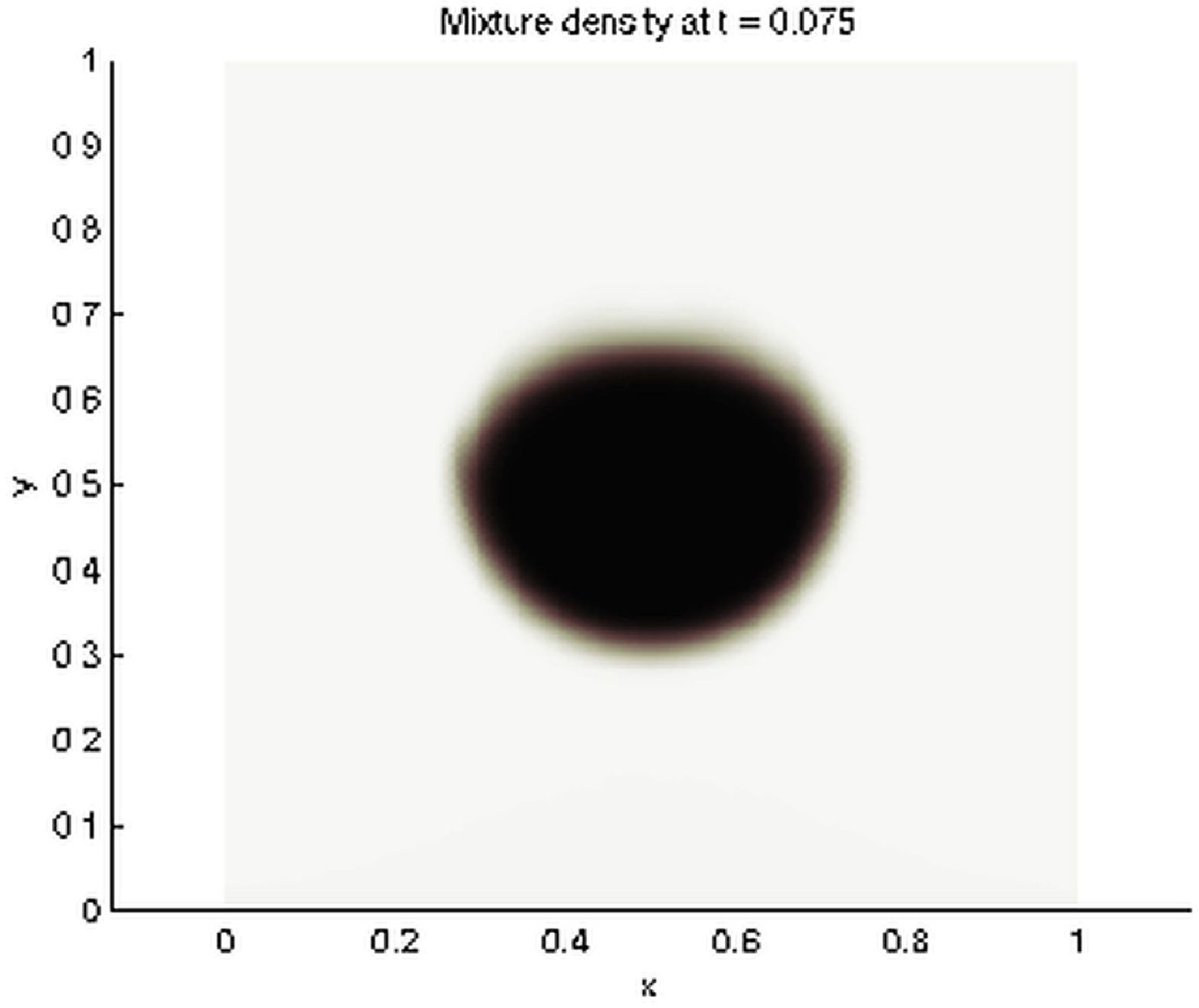}}
	\caption[Initial configuration and the beginning of the fall]{Water drop test case. Initial configuration and the beginning of the fall.}
	\label{fig:debutDrop}
\end{figure}

\begin{figure}
	\centering
	\subfigure[$t = 0.325$ s]%
	{\includegraphics[width=0.46\textwidth]{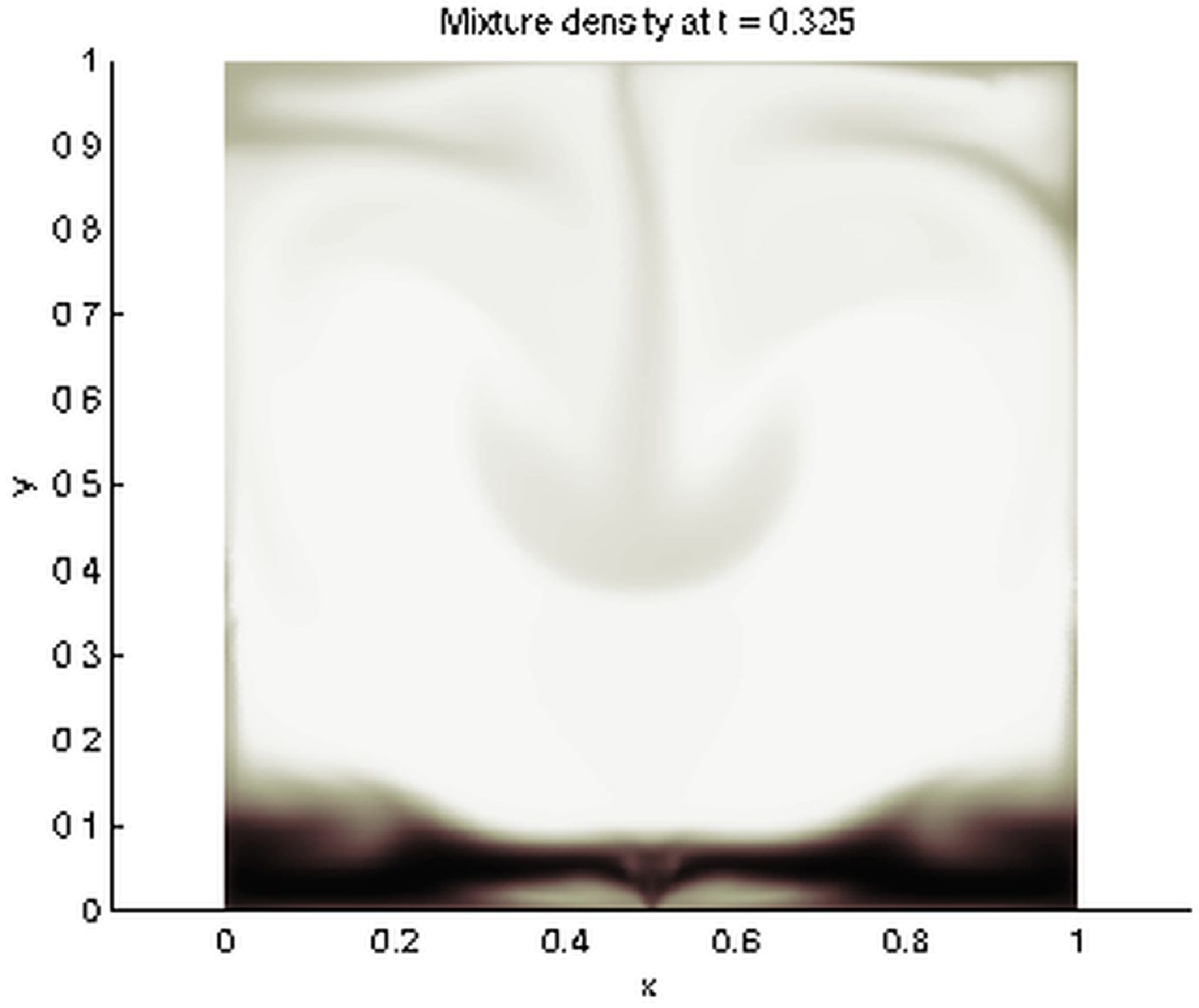}} \quad
	\subfigure[$t = 0.35$ s]%
	{\includegraphics[width=0.46\textwidth]{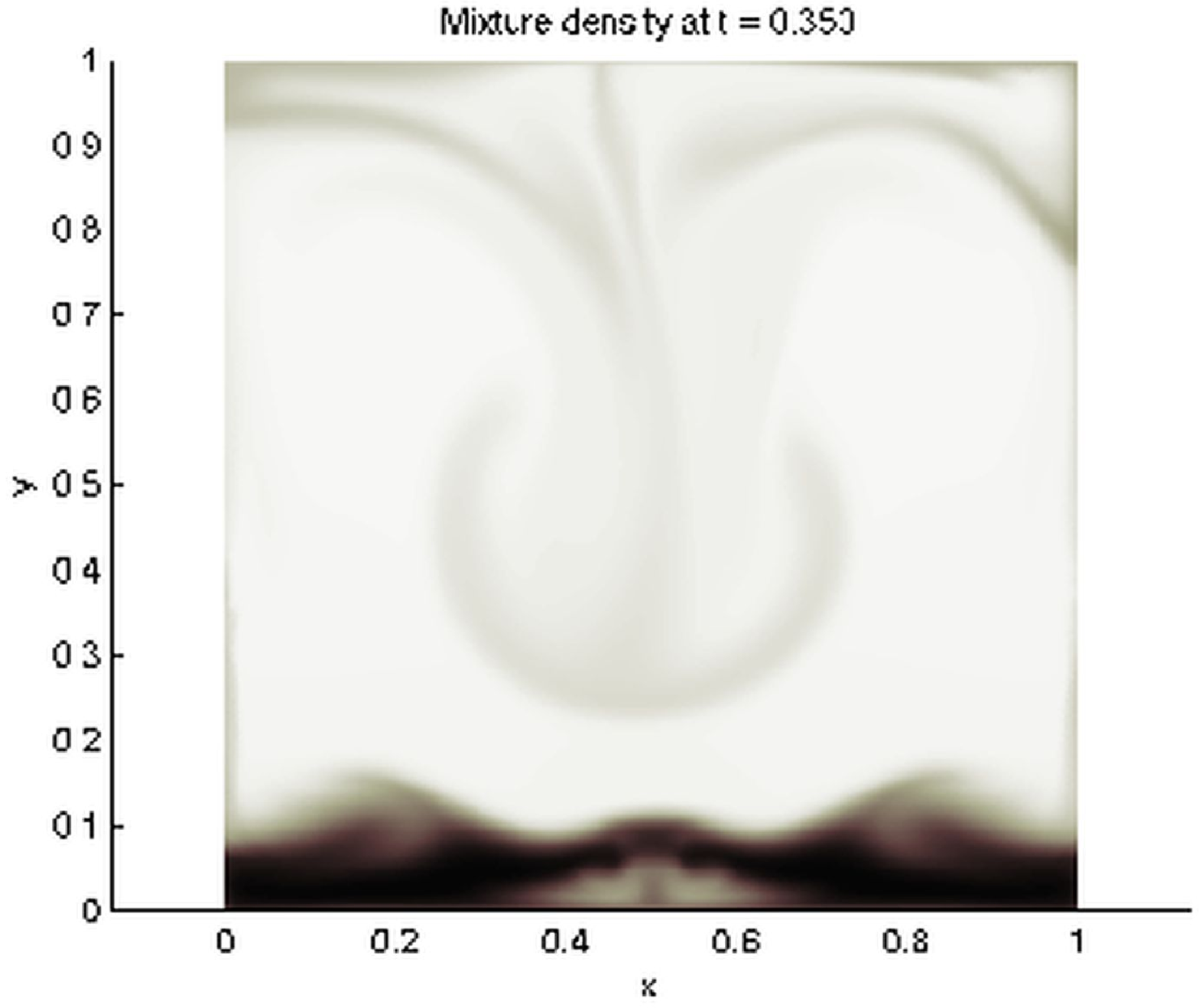}}
	\caption[Water drop test case. Side jets interflow at the center]{Water drop test case. Side jets flowing down the centerline.}
	\label{fig:DropAsym}
\end{figure}

\begin{figure}
	\centering
	\subfigure[$t = 0.4$ s]%
	{\includegraphics[width=0.46\textwidth]{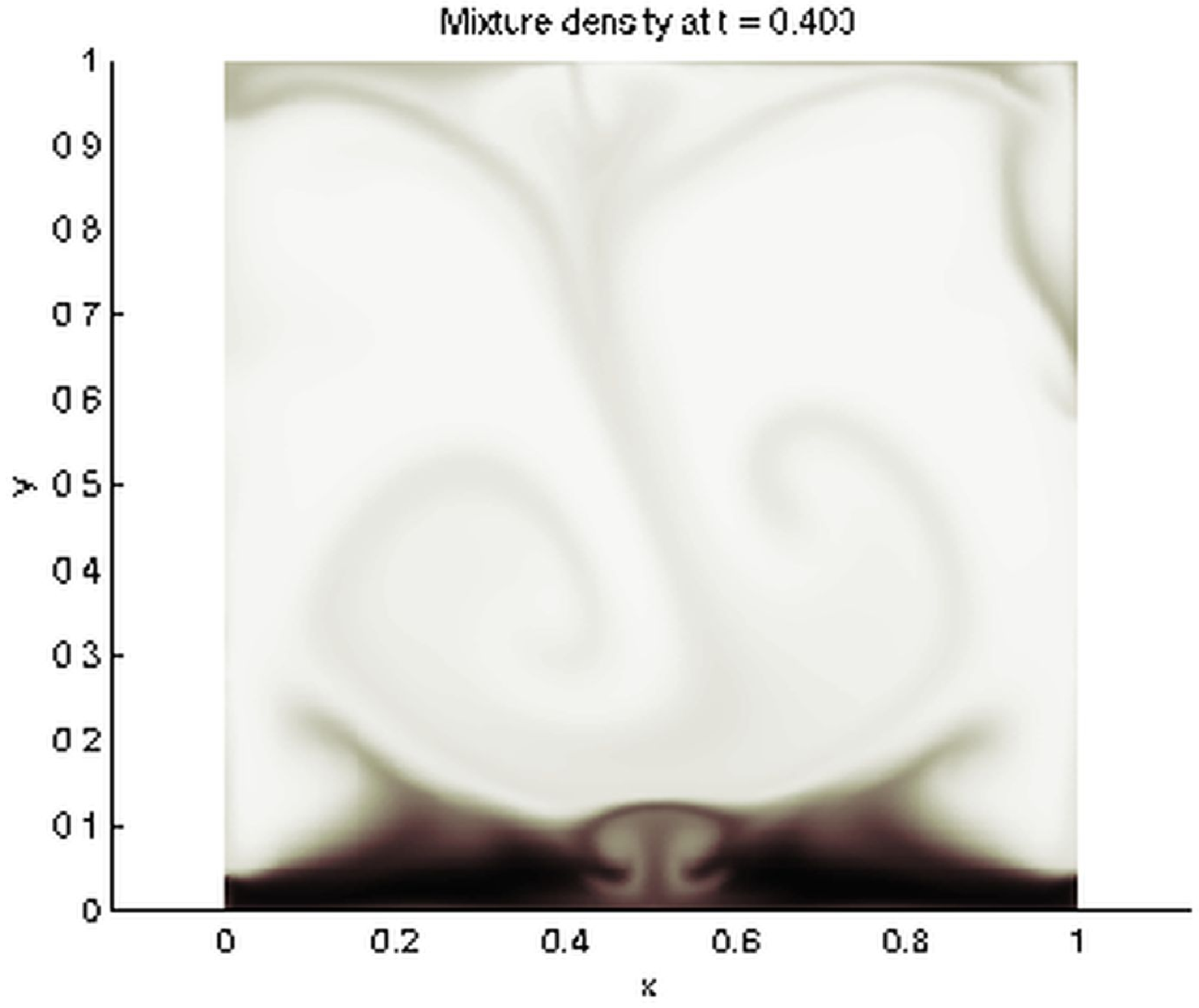}} \quad
	\subfigure[$t = 0.45$ s]%
	{\includegraphics[width=0.46\textwidth]{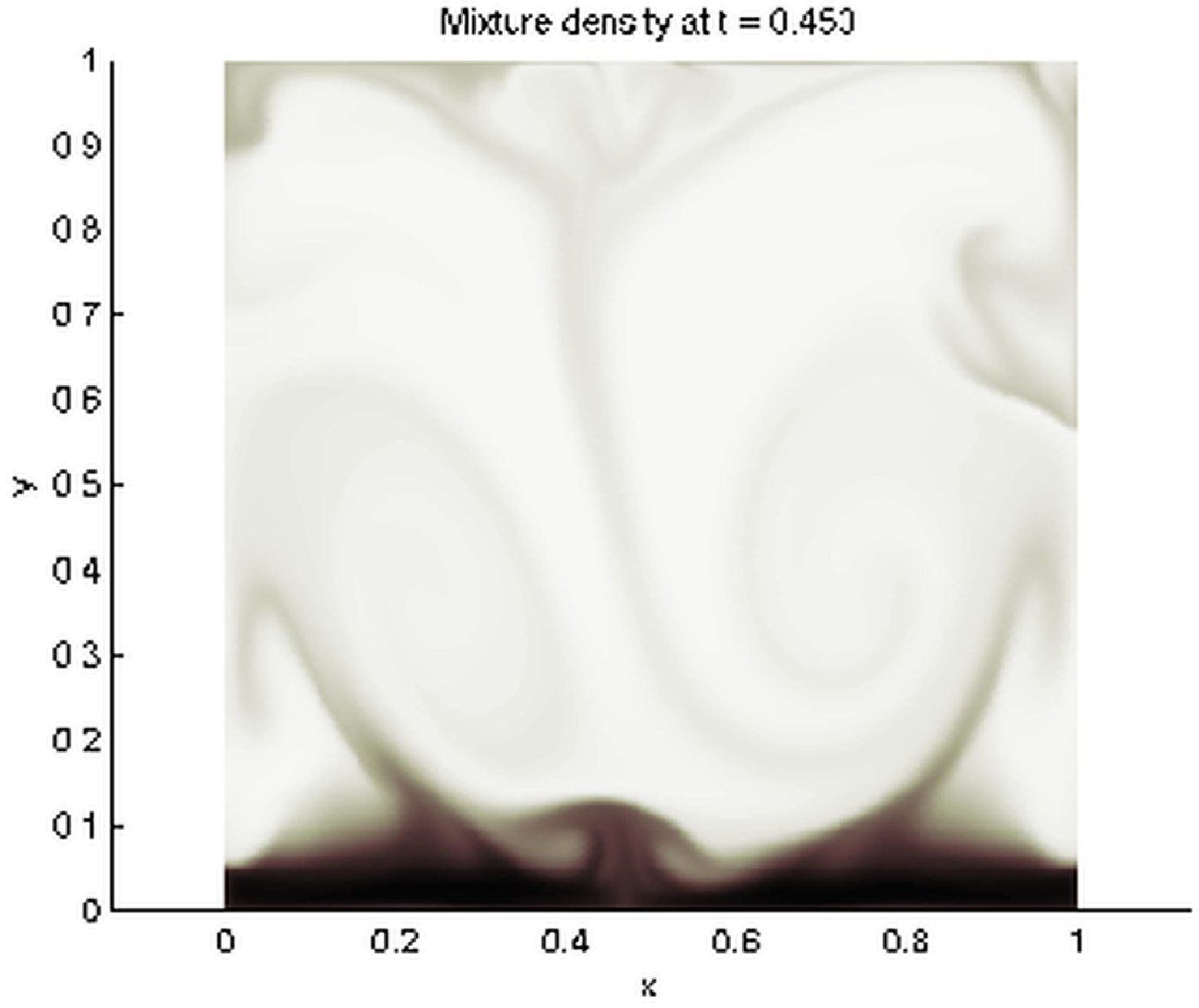}}
	\caption[Central jet reflection from the bottom]{Water drop test case. Central jet reflection from the bottom.}
	\label{fig:lastDrop}
\end{figure}

\begin{figure}
	\centering
		\includegraphics[width=0.7\textwidth]{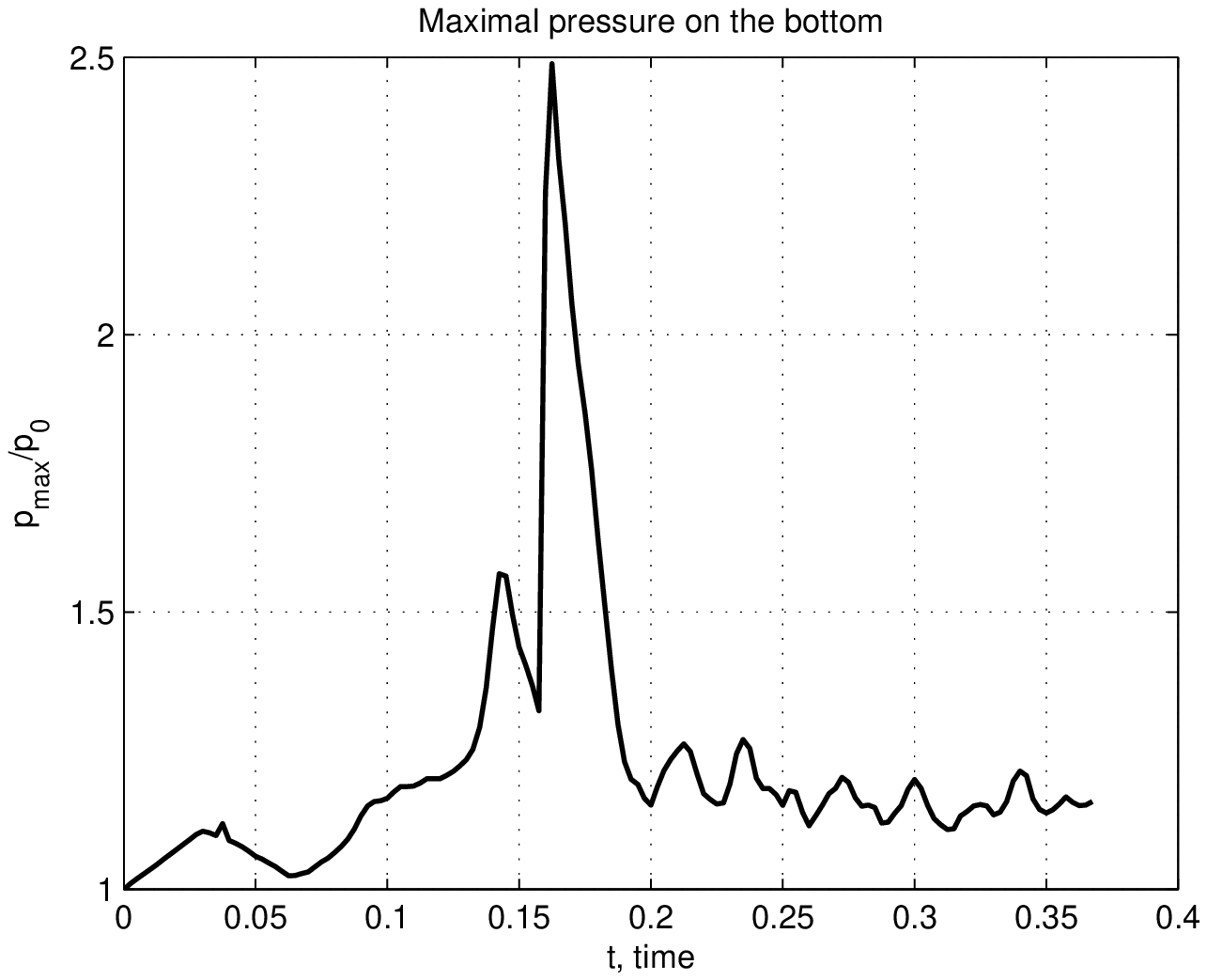}
	\caption{Water drop test case. Maximum bottom pressure as a function of time.}
	\label{fig:bottompress}
\end{figure}

The geometry and initial condition for this test case are shown on \figurename~\ref{fig:drop_water}. Initially the velocity field is taken to be zero. The values of the other parameters are given in Table \ref{tab:diphaseparams}. The mesh used in this computation contained about $92 000$ control volumes (again they were triangles). The results of this simulation are presented in Figures \ref{fig:debutDrop}--\ref{fig:lastDrop}. In \figurename~\ref{fig:bottompress} we plot the maximal pressure on the bottom as a function of time:
\begin{equation*}
  t \longmapsto \max_{(x,y) \in [0,1]\times 0} p(x,y,t).
\end{equation*}
The pressure exerted on the bottom reaches $2.5p_0$ due to the drop impact at $t\approx 0.16$ s.

\begin{figure}
	\centering
	\subfigure[$t = 0.1$ s]%
	{\includegraphics[width=0.46\textwidth]{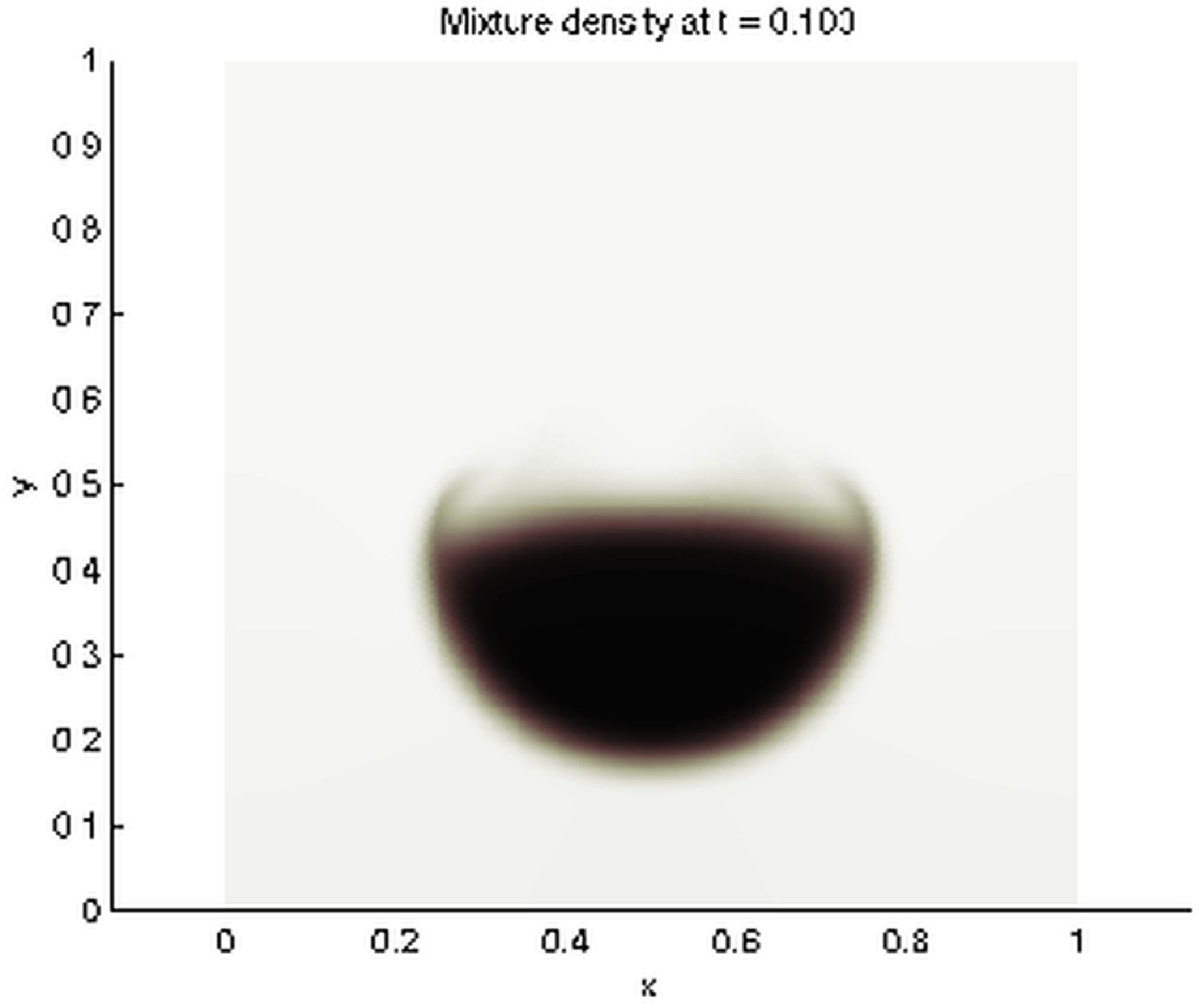}} \quad
	\subfigure[$t = 0.125$ s]%
	{\includegraphics[width=0.46\textwidth]{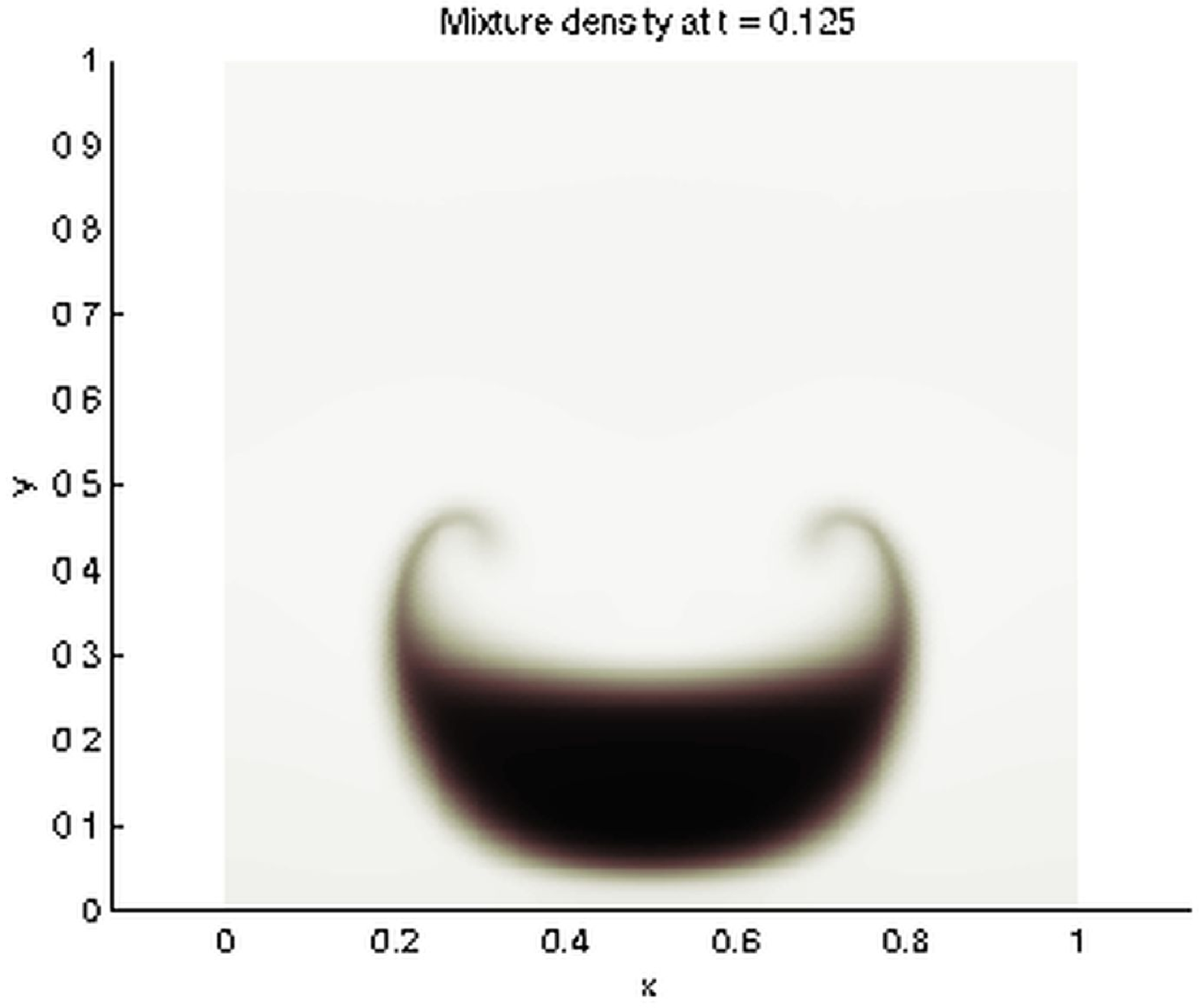}}
	\caption[Drop approaching container bottom]{Water drop test case. Drop approaching the bottom of the container.}
\end{figure}

\begin{figure}
	\centering
	\subfigure[$t = 0.135$ s]%
	{\includegraphics[width=0.46\textwidth]{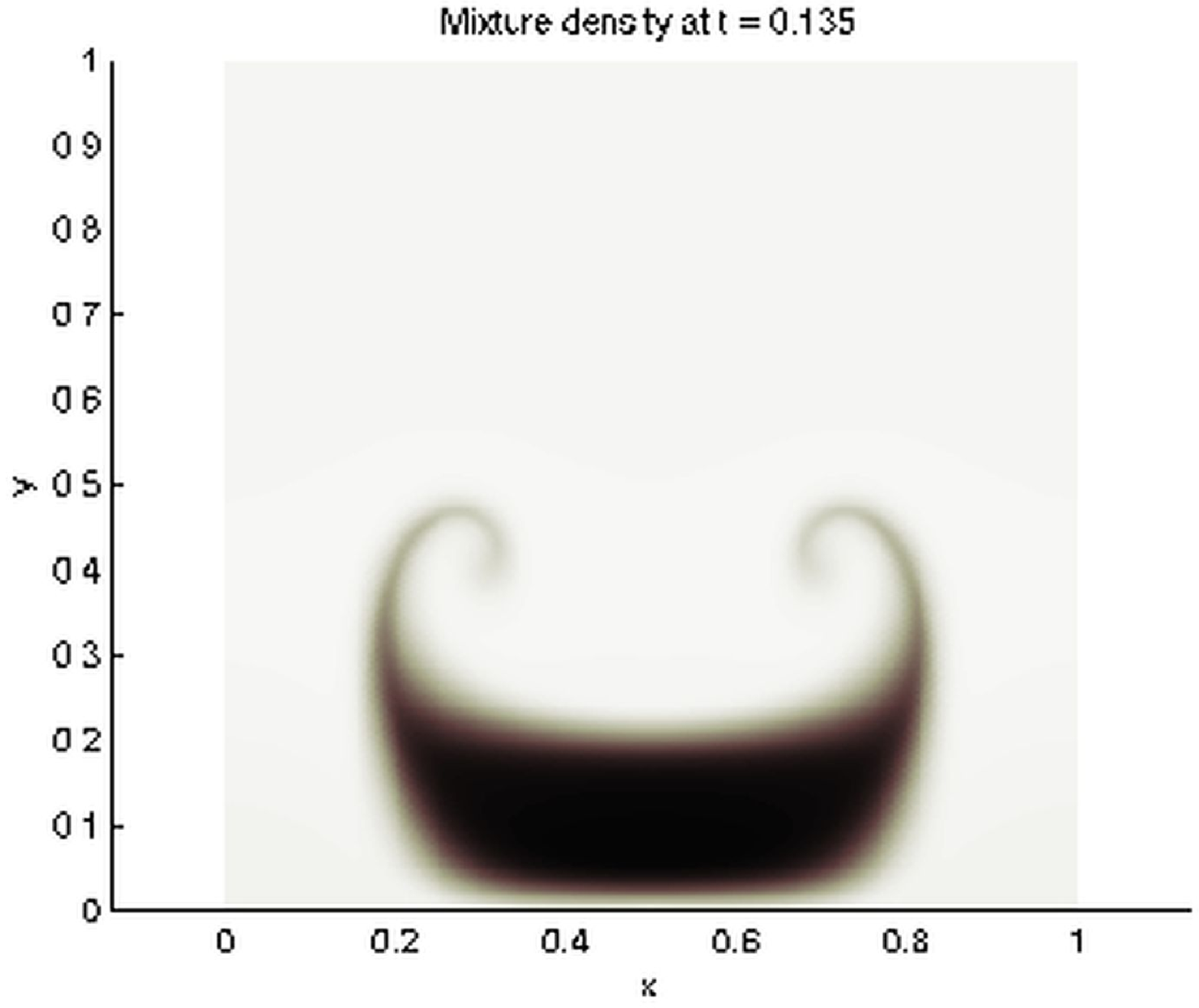}} \quad
	\subfigure[$t = 0.15$ s]%
	{\includegraphics[width=0.46\textwidth]{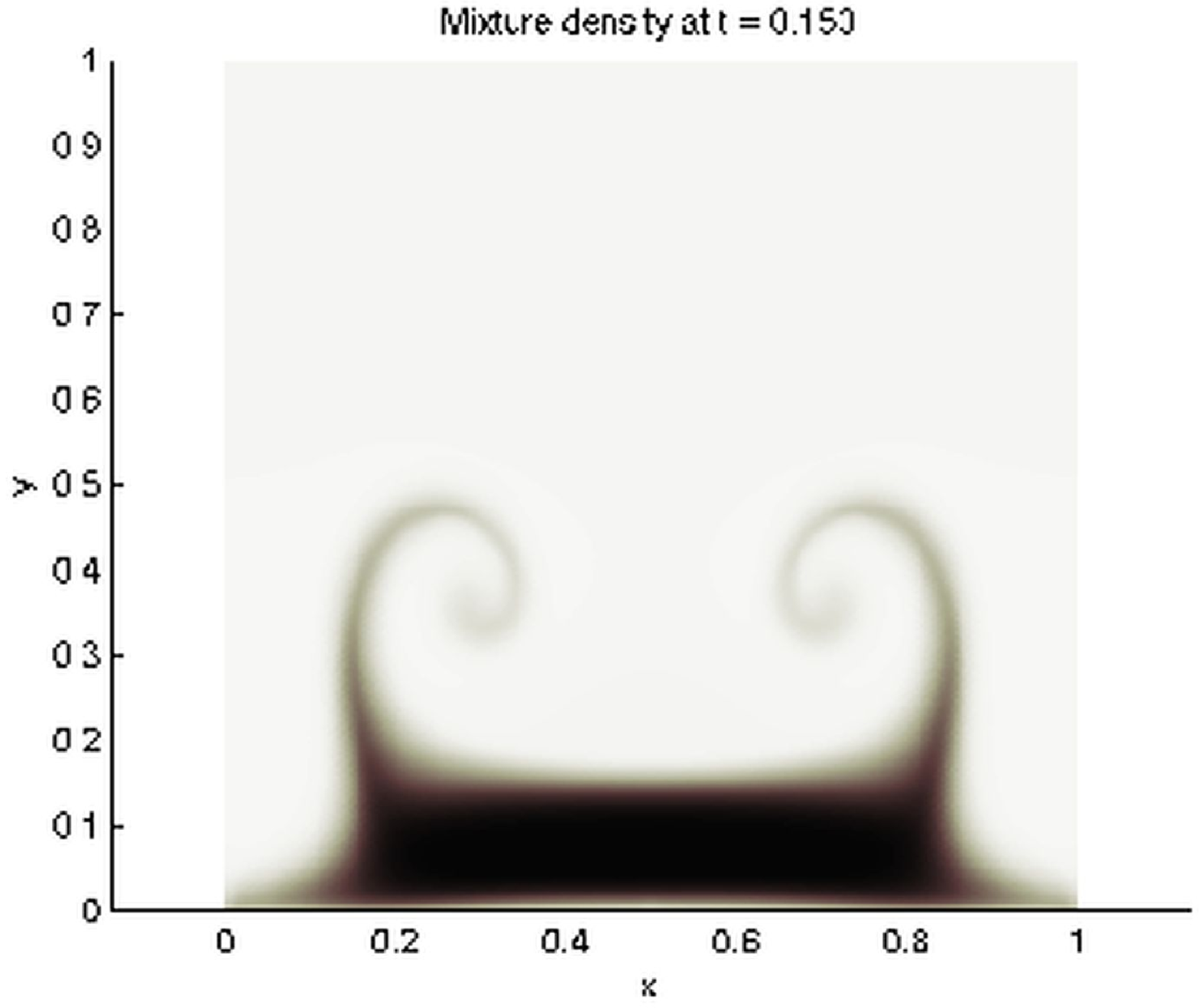}}
	\caption[Drop/bottom compressible interaction]{Water drop test case. Drop/bottom compressible interaction.}
\end{figure}

\begin{figure}
	\centering
	\subfigure[$t = 0.175$ s]%
	{\includegraphics[width=0.46\textwidth]{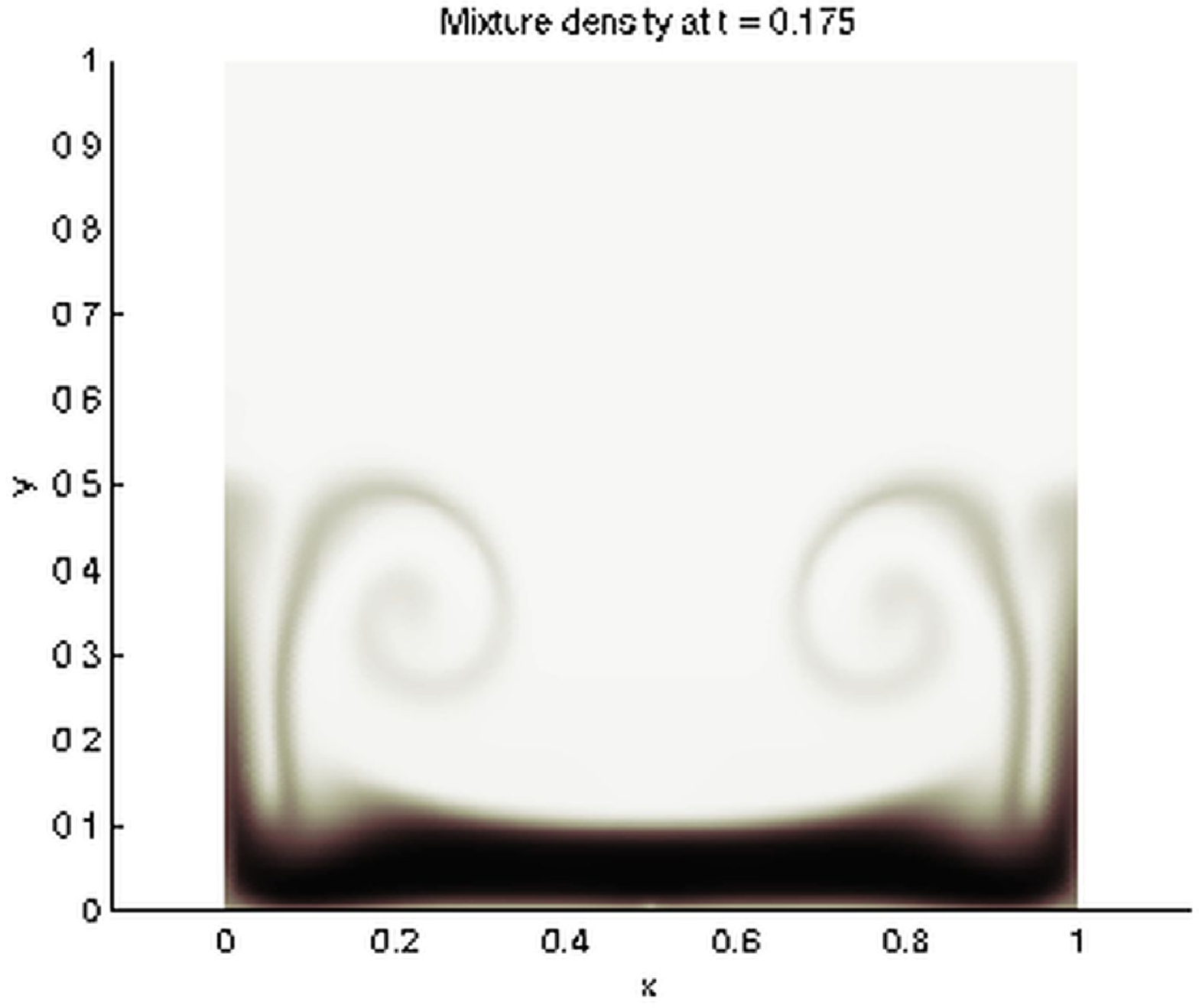}} \quad
	\subfigure[$t = 0.2$ s]%
	{\includegraphics[width=0.46\textwidth]{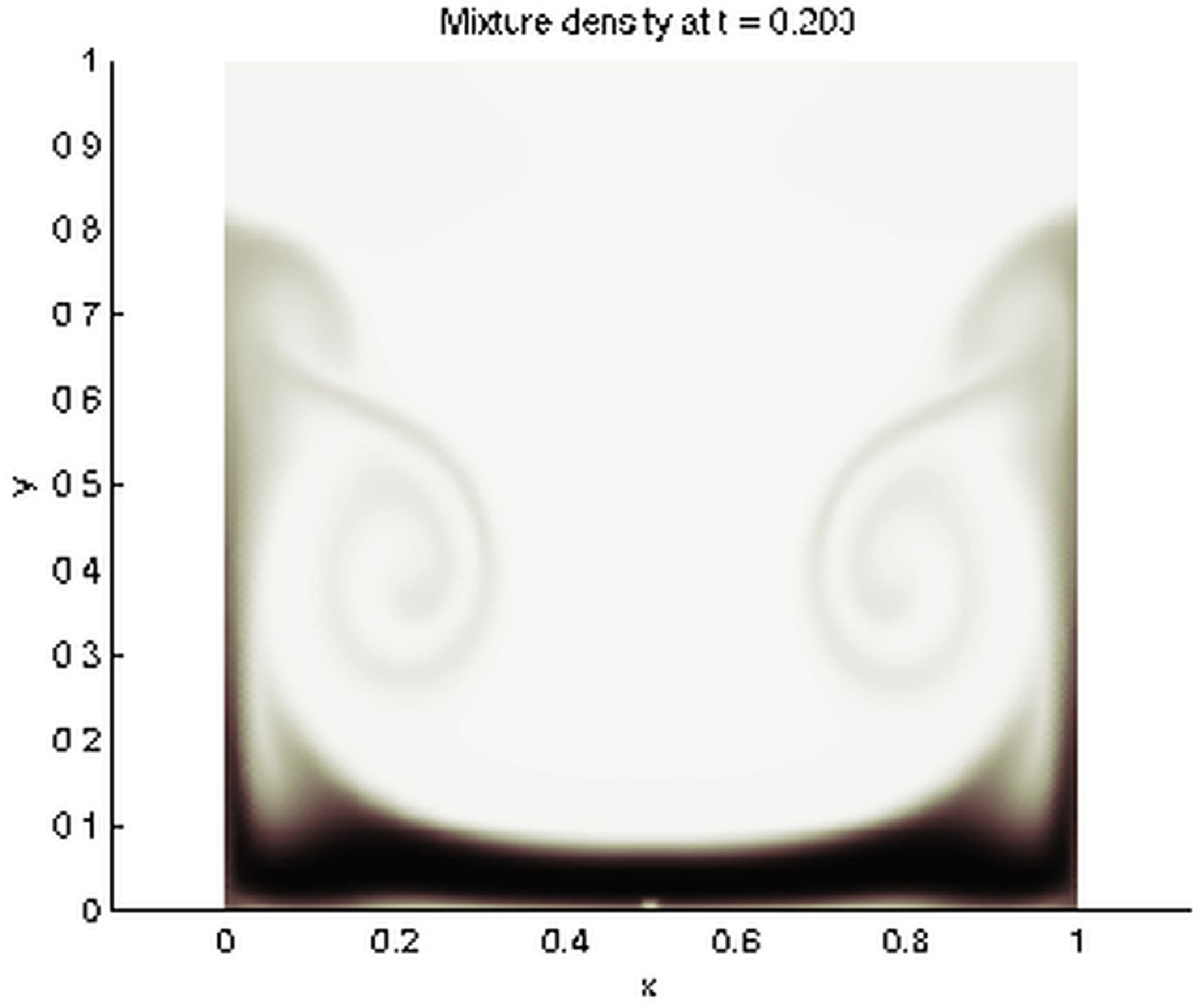}}
	\caption[Water drop test case. Formation of vertical jets.]{Water drop test case. Formation of vertical jets.}
\end{figure}

\begin{figure}
	\centering
	\subfigure[$t = 0.225$ s]%
	{\includegraphics[width=0.46\textwidth]{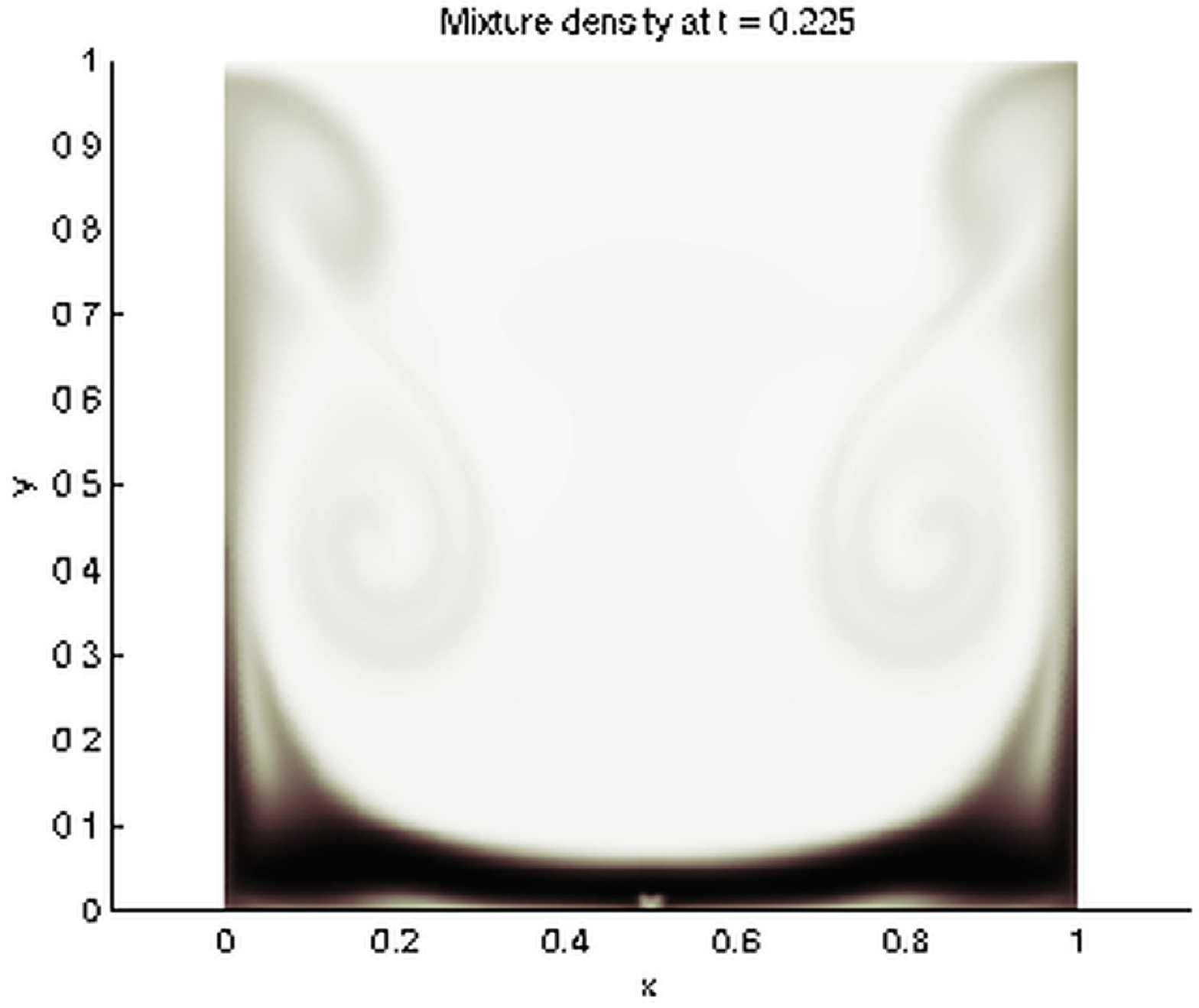}} \quad
	\subfigure[$t = 0.275$ s]%
	{\includegraphics[width=0.46\textwidth]{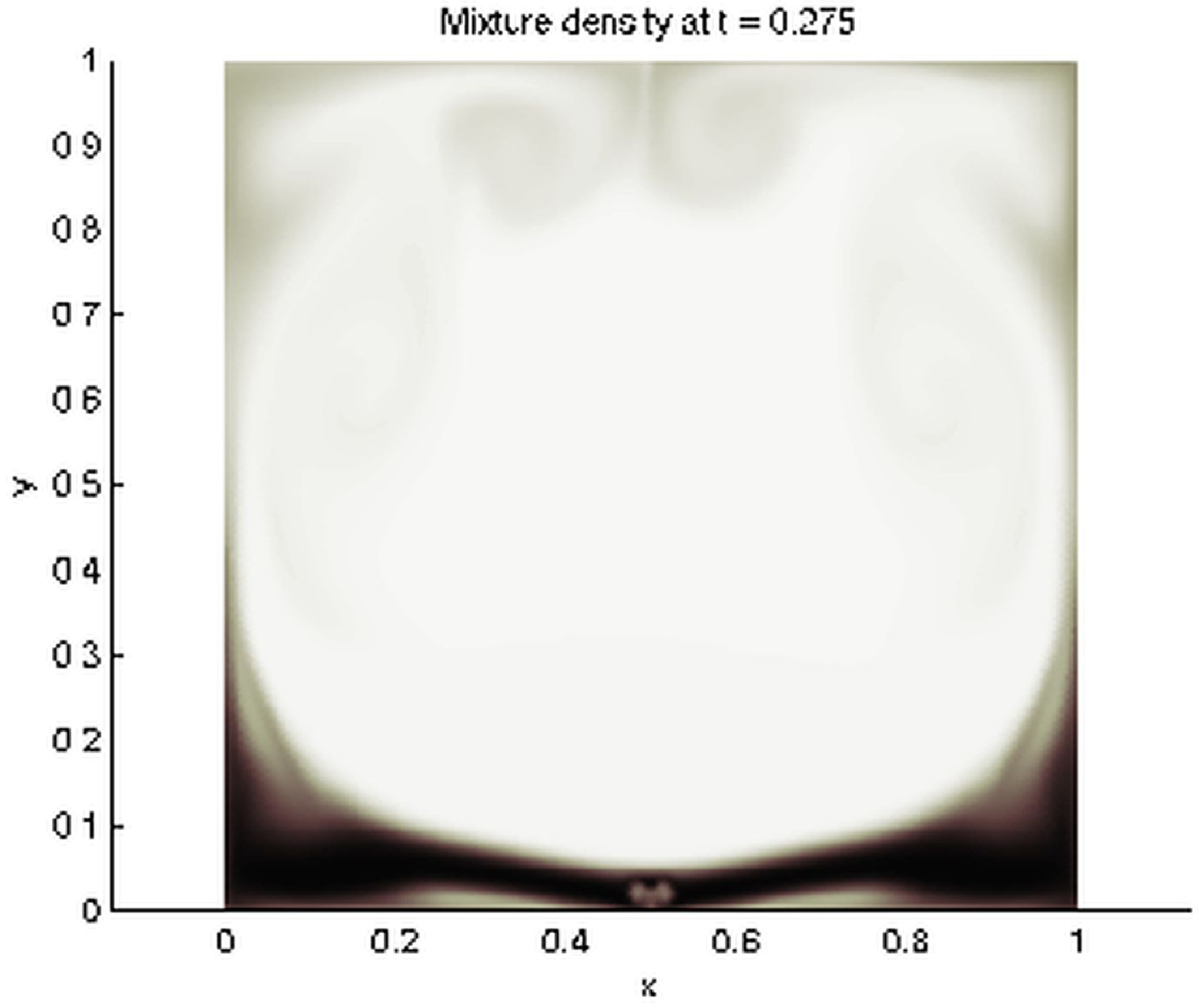}}
	\caption[Water drop test case. Side jets crossing]{Water drop test case. Crossing of side jets.}
\end{figure}

\section{Conclusions}

In this article we have presented a simple mathematical model for simulating water wave impacts. Associated to this model, which avoids the costly capture of free surfaces, we have built a numerical solver which is: (i)~ second-order accurate on smooth solutions, (ii)~stable even for solutions with very strong gradients (and solutions with shocks) and (iii)~locally {\it exactly} conservative with respect to the mass of each fluid, momentum and total energy. This last property, (iii), which is certainly the most desirable from the physical point of view, is an immediate byproduct of our cell-centered finite volume method.

We have shown here the good behavior of this framework on simple test cases and we are presently working on quantitative comparisons in the context of real applications.

\section*{Acknowledgments}
This work has been partially supported by ANR HEXECO, Project n$^o$ BLAN$07-1\_192661$, and by the 2008 Framework Program for Research,
Technological development and Innovation of the Cyprus Research Promotion Foundation under the Project A$\Sigma$TI/0308(BE)/05. 
The second author acknowledges support through a PhD grant from the French Ministry of Research.

\bibliographystyle{plain}
\bibliography{biblio}

\begin{thebibliography}{10}

\bibitem{Bagnold1939}
R.A. Bagnold.
\newblock Interim report on wave pressure research.
\newblock {\em Proc. Inst. Civil Eng.}, 12:201--26, 1939.

\bibitem{Bredmose2005}
H.~Bredmose.
\newblock Flair: A finite volume solver for aerated flows.
\newblock Technical report, 2005.

\bibitem{Bredmose1}
H.~Bredmose, D.~H. Peregrine, G.~N. Bullock, C.~C.~Obhrai, G.~Müller, and
  G.~Wolters.
\newblock Extreme wave impact pressures and the effect of aeration.
\newblock In {\em Int. Workshop on Water Waves and Floating Bodies, Cortona,
  Italy}, 2004.

\bibitem{Bullock2007}
G.~N. Bullock, C.~Obhrai, D.~H. Peregrine, and H.~Bredmose.
\newblock Violent breaking wave impacts. part 1: Results from large-scale
  regular wave tests on vertical and sloping walls.
\newblock {\em Coastal Engineering}, 54:602--617, 2007.

\bibitem{Bullock2001}
G.N. Bullock, A.R. Crawford, P.J. Hewson, M.J.A. Walkden, and P.A.D. Bird.
\newblock The influence of air and scale on wave impact pressures.
\newblock {\em Coastal Engineering}, 42:291--312, 2001.

\bibitem{Cole1948}
R.H. Cole.
\newblock {\em Underwater explosions}.
\newblock Princeton University Press, 1948.

\bibitem{DDG2008}
F.~Dias, D.~Dutykh, and J.-M. Ghidaglia.
\newblock A compressible two-fluid model for the finite volume simulation of
  violent aerated flows. {A}nalytical properties and numerical results.
\newblock {\em http://hal.archives-ouvertes.fr/hal-00279671/}, pages 1--38,
  2008.

\bibitem{Dutykh2007a}
D.~Dutykh.
\newblock {\em Mathematical modelling of tsunami waves}.
\newblock PhD thesis, \'{E}cole {N}ormale {S}up\'{e}rieure de {C}achan, 2007.

\bibitem{Ghidaglia2001}
J.-M. Ghidaglia, A.~Kumbaro, and G.~Le Coq.
\newblock On the numerical solution to two fluid models via cell centered
  finite volume method.
\newblock {\em Eur. J. Mech. B/Fluids}, 20:841--867, 2001.

\bibitem{Ghidaglia2005}
J.-M. Ghidaglia and F.~Pascal.
\newblock The normal flux method at the boundary for multidimensional finite
  volume approximations in cfd.
\newblock {\em European Journal of Mechanics B/Fluids}, 24:1--17, 2005.

\bibitem{Godunov1979}
S.K. Godunov, A.~Zabrodine, M.~Ivanov, A.~Kraiko, and G.~Prokopov.
\newblock {\em R\'{e}solution num\'{e}rique des probl\`{e}mes
  multidimensionnels de la dynamique des gaz}.
\newblock Editions Mir, Moscow, 1979.

\bibitem{Helluy2005}
P.~Helluy, F.~Golay, J.-P. Caltagirone, P.~Lubin, S.~Vincent, D.~Drevard,
  R.~Marcer, P.~Fraunie, N.~Seguin, S.~Grilli, A.-C. Lesage, A.~Dervieux, and
  O.~Allain.
\newblock {N}umerical simulation of wave breaking.
\newblock {\em {M}athematical {M}odelling and {N}umerical {A}nalysis},
  39(3):591--607, 2005.

\bibitem{Ishii1975}
M.~Ishii.
\newblock {\em Thermo-Fluid Dynamic Theory of Two-Phase Flow}.
\newblock Eyrolles, Paris, 1975.

\bibitem{James2001}
G.~James.
\newblock Internal travelling waves in the limit of a discontinuously
  stratified fluid.
\newblock {\em Arch. Rational Mech. Anal.}, 160:41--90, 2001.

\bibitem{Mehaute1995}
B.~Le Mehauté and S.~Wang.
\newblock {\em Water Waves Generated by Underwater Explosion, Advanced Series
  on Ocean Engineering, Vol. 10}.
\newblock World Scientific, Singapore, 1995.

\bibitem{Bredmose3}
D.~H. Peregrine, H.~Bredmose, G.~Bullock, A.~Hunt, and C.~Obhrai.
\newblock Water wave impact on walls and the role of air.
\newblock In {\em Proc. 30th Int. Conf. Coast. Engng., San Diego (ed. J. M.
  Smith), vol. 5, pp. 4494-4506. ASCE}, 2006.

\bibitem{Bredmose2}
D.~H. Peregrine, H.~Bredmose, G.~Bullock, C.~Obhrai, G.~Müller, and G.~Wolters.
\newblock Water wave impact on walls and the role of air.
\newblock In {\em Proceedings of the 29th International Conference on Coastal
  Engineering, Lisbon 2004, vol. 4, pp. 4005-4017. ASCE}, 2004.

\bibitem{Peregrine1996}
D.H. Peregrine and L.~Thais.
\newblock The effect of entrained air in violent water impacts.
\newblock {\em J. Fluid Mech.}, 325:377--97, 1996.

\bibitem{leer2006}
B.~van Leer.
\newblock Upwind and high-resolution methods for compressible flow: From donor
  cell to residual-distribution schemes.
\newblock {\em Communications in Computational Physics}, 1:192--206, 2006.

\bibitem{Whitham1999}
G.B. Whitham.
\newblock {\em Linear and nonlinear waves}.
\newblock John Wiley \& Sons Inc., New York, 1999.

\bibitem{Wood2000}
D.J. Wood, D.H. Peregrine, and T.~Bruce.
\newblock Wave impact on wall using pressure-impulse theory. i. trapped air.
\newblock {\em Journal of Waterway, Port, Coastal and Ocean Engineering},
  126(4):182--190, 2000.

\end{thebibliography}
\end{document}